\documentclass[apj,iop]{emulateapj}
\usepackage{latexsym,graphicx,rotating,amsmath, epsfig, natbib}
\usepackage{apjfonts}
\renewcommand{\vec}{\boldsymbol}
\newcommand{\sol}{\odot}
\newcommand{\del}{\nabla}

\newcommand{\scrD}{\mathcal{D}}
\newcommand{\scrQ}{\mathcal{Q}}
\newcommand{\scrO}{\mathcal{O}}

\newcommand{\ms}{\mathrm{m}\thinspace \mathrm{s}^{-1}}
\newcommand{\advbm}{\left[ \langle B_r \rangle \frac{\partial}{\partial r}+\frac{ \langle B_{\theta} \rangle }{r}\frac{\partial}{\partial \theta}\right] }

\newcommand\jfm{\rmfamily{J. Fluid Mech.}}

\shorttitle{Magnetic Cycles in a Solar-type Star}
\shortauthors{Brown, Miesch, Browning, Brun \& Toomre}

\slugcomment{Accepted for publication in ApJ}

\begin{document}

  \title{Magnetic Cycles in a Convective Dynamo Simulation of a Young Solar-type Star}

  \author{Benjamin P.\ Brown}
  \affil{Dept.\ Astronomy, University of Wisconsin, Madison, WI 53706-1582}
  \affil{Center for Magnetic Self Organization in Laboratory and Astrophysical Plasmas, 
         University of Wisconsin, 1150 University Avenue, Madison, WI 53706}
  \affil{JILA and Dept.\ Astrophysical \& Planetary Sciences, University of Colorado, Boulder, CO 80309-0440}
  \email{bpbrown@astro.wisc.edu}
  \author{Mark S.\ Miesch}
  \affil{High Altitude Observatory, NCAR, Boulder, CO 80307-3000}
  \author{Matthew K.\ Browning}
  \affil{Canadian Institute for Theoretical Astrophysics, University of Toronto, Toronto, ON M5S3H8 Canada}
  \author{Allan Sacha Brun}
  \affil{Laboratoire AIM Paris-Saclay, CEA/Irfu Universit\'e Paris-Diderot CNRS/INSU, 91191 Gif-sur-Yvette Cedex, France}
  \and
  \author{Juri Toomre}
  \affil{JILA and Dept.\ Astrophysical \& Planetary Sciences, University of Colorado, Boulder, CO 80309-0440}

  \begin{abstract}
   Young solar-type stars rotate rapidly and many are
   magnetically active.  Some appear to undergo magnetic cycles
   similar to the 22-year solar activity cycle.  We conduct
   simulations of dynamo action in rapidly rotating suns with the
   3D MHD anelastic spherical harmonic
   (ASH) code to explore dynamo action achieved in
   the convective envelope of a solar-type star rotating at 5 times the
   current solar rotation rate.  We find that dynamo action builds
   substantial organized global-scale magnetic fields in the midst of the convection zone.
   Striking magnetic wreaths span the convection zone and coexist with
   the turbulent convection.  A surprising feature of this
   wreath-building dynamo is its rich time-dependence.  The dynamo exhibits
   cyclic activity and undergoes quasi-periodic 
   polarity reversals where both the global-scale poloidal and
   toroidal fields change in sense on a roughly 1500~day time scale.
   These magnetic activity patterns emerge spontaneously from the
   turbulent flow and are more organized temporally and spatially than those
   realized in our previous simulations of the solar dynamo.
   We assess in detail the competing processes of magnetic field
   creation and destruction within our simulations that contribute to
   the global-scale reversals.  We find that the mean toroidal fields
   are built primarily through an $\Omega$-effect, while the mean
   poloidal fields are built by turbulent correlations which are not
   necessarily well represented by a simple $\alpha$-effect.  During a
   reversal the magnetic wreaths propagate towards the polar regions,
   and this appears to arise from a poleward propagating dynamo wave.
   As the magnetic fields wax and wane in strength and flip in
   polarity, the primary response in the convective flows involves the
   axisymmetric differential rotation which varies on similar
   time scales.  Bands of relatively fast and slow fluid propagate
   towards the poles on time scales of roughly 500~days and are
   associated with the magnetic structures that propagate in the same
   fashion.  In the Sun, similar patterns are observed in the poleward
   branch of the torsional oscillations, and these may represent
   poleward propagating magnetic fields deep below the solar surface.
 \end{abstract}
  \keywords{convection -- MHD -- stars:interiors --
  stars:rotation -- stars: magnetic fields -- Sun:interior}


\section{Introduction}\label{intro}

Motivated by the rich observational landscape of stellar magnetism and
by our own Sun's cycles of activity, we have undertaken 3D simulations
of convection and dynamo action in solar type stars.  These
simulations have explored how global-scale flows of differential
rotation and meridional circulation are established in rapidly
rotating suns \citep{Brown_et_al_2008} and how dynamo action might be
realized in these stars.  

In \mbox{Paper I} \citep{Brown_et_al_2010a}
we studied dynamo action in a star rotating three times faster than
our current Sun and found that global-scale magnetic structures can arise
naturally in the midst of the stellar convection zone.  These wreaths
of magnetism were stable for long periods of time, maintaining their
identity for many thousands of days, and did not require a stably
stratified tachocline to survive the turbulent convection.  Rather they
coexisted with it.

In this paper we explore a convective dynamo simulation in a
solar-like star rotating five times the current solar rate.  
Here, global-scale magnetic wreaths still form in the convection zone but
now they become time dependent and undergo repeated cycles of magnetic
polarity reversal.  These wreaths are significantly more complex than
the steady wreaths discussed in \mbox{Paper~I}.

\subsection{Observational Landscape}

Magnetism is a nearly ubiquitous feature of solar-type stars.  Young, rapidly
rotating suns appear to have much stronger magnetic fields at their
surfaces.  Observations reveal a clear correlation 
between rotation and magnetic activity, as inferred from proxies
such as X-ray and chromospheric emission
\citep{Schrijver&Zwaan_2000,Pizzolato_et_al_2003}, but the
observational landscape is complex with few other well-established
trends to constrain dynamo models \citep[e.g.,][]{Rempel_2008,Lanza_2010}.

A fundamental characteristic of the solar dynamo is its regular cycles
of sunspots.  Indeed, the 22-year solar activity cycle stands out as
one of the most remarkable and enigmatic examples of magnetic
self-organization in nature.  
Solar type stars appear to undergo magnetic cycles 
with periods ranging from several years to several decades.
Most observational investigations of stellar activity cycles
rely on long-term monitoring of photospheric, chromospheric,
or coronal emission using the Sun as a baseline for calibration 
and comparison \citep[e.g.,][]{Baliunas_et_al_1995,Hempelmann_et_al_1996,
Messina&Guinan_2002,Hall_2008,Olah_et_al_2009,Lanza_2010}.  
Photometric and spectroscopic variability are linked to magnetic
activity and periodic modulation on time scales of years to
decades is interpreted as an activity cycle.  The most 
extensive such survey is still the Mount Wilson 
HK Project which monitored chromospheric emission
in 111 solar-type stars from 1966--1991
\citep{Baliunas_et_al_1995}.  About half of the stars in the sample (51) 
show clear signs of cyclic activity, including 21 with well-defined 
cycle periods ranging between 7 and 25 years.  The 
remaining stars exhibit either irregular variability
with no clear systematic variation (29) or
smooth time series with flat or linear trends (31).
Longer-term but more sporadic photometric measurements
are available for a few dozen stars and these, along
with the Mount Wilson data, show evidence for 
multiple periodicities and possible variations 
in the apparent starspot cycle periods over 
the course of multiple decades \citep{Olah_et_al_2009}.
More rapidly rotating stars generally have shorter cycles
\citep{Saar&Brandenburg_1999}, but robust scaling relationships with
stellar mass, rotation rate and other fundamental parameters have
remained elusive.  The shortest stellar activity cycles observed to
date have periods of roughly 1 to 1.6 years, and occur in rapidly
rotating F-type stars \citep{Metcalfe_et_al_2010, Garcia_et_al_2010}.

\subsection{Elements of Cyclic Activity}

The magnetic fields observed at the surfaces of late-type stars must arise from
dynamo action in the stellar convection zones, as in the solar dynamo. 
Self-organization in turbulent dynamos is intimately associated
with helicity and shear.  In many mean-field dynamo theories, poloidal
magnetic fields are built by turbulent correlations in the convection
through what is known as an $\alpha$-effect, and these correlations
are enhanced if the flow is helical.  Rotation and stratification impart helicity, 
both kinetic and magnetic, and can also lead to large-scale shearing
flows of differential rotation.  Global-scale shear promotes self-organization
through a process known as the $\Omega$-effect where mean toroidal
fields are generated from poloidal fields.  Dynamos employing these
two effects are known as $\alpha$-$\Omega$ dynamos, and these models
form the basis of much of our theoretical understanding of cyclic
dynamos is stars like the sun \citep[e.g.,][]{Charbonneau_2010}.  

The central role of shear and helicity in establishing cyclic activity
has been confirmed by global numerical simulations of convective
dynamos.  The first 3D magnetohydrodynamic (MHD)
simulations of convection in rotating spherical shells to exhibit
cyclic behavior were explored in detail by \citet{Gilman_1983}.
Recent simulations of global convective dynamos have focused on
rapidly-rotating Boussinesq systems in deep convective shells, often
in the context of planetary dynamos
\citep[reviewed by][]{Busse_2000,Roberts&Glatzmaier_2001,Christensen&Aubert_2006}.
Here the magnetic field is often dominated by the axisymmetric dipole
component and is typically either stable in time or undergoes chaotic
reversals.  In planetary dynamos, rapid rotation, deep shell
geometries, and minimal density stratifications promote quasi-2D
convective columns that are strongly aligned with the rotation axis.
Stellar convection occurs under quite different conditions.

In the solar convection zone, density stratification
is of fundamental importance and the rotational 
influence is moderate (Rossby number $\mathrm{Ro} \sim 1$), giving rise
to intricate, three-dimensional, highly turbulent 
convective structures spanning many spatial and 
temporal scales \citep{Miesch_et_al_2008}.  Dynamo simulations
in this parameter regime produce complex magnetic
topologies, with more than 95\% of the magnetic
energy in the fluctuating (non-axisymmetric) field
components \citep{Brun_et_al_2004}.  Mean magnetic 
fields are likewise complex, with multipolar structure and transient
toroidal ribbons and sheets.  Polarity reversals
of the dipole component occur but they are 
irregular in time.  

The presence of an overshoot region below the convection zone and a tachocline 
of rotational shear there promotes mean-field generation, 
producing persistent bands of toroidal flux antisymmetric 
about the equator while strengthening and stabilizing the dipole 
moment \citep{Browning_et_al_2006,Browning_et_al_2007,Miesch_et_al_2009}.
However, these simulations did not exhibit systematic magnetic cycles. 
Recent results by \citep{Ghizaru_et_al_2010} achieve both large-scale 
organization and cyclic reversals of polarity in a convection simulation
with an underlying tachocline.   In that simulation, substantial mean
magnetic fields are present in both the tachocline and the bulk of the
convection zone.  The simulation spans roughly 93,000 days
(225~years) with polarity reversals occurring on roughly 11,000 day
(30~year) periods.  Convective simulations in spherical wedge
geometries with self-consistently established differential rotation
profiles and imposed tachoclines also achieve organized fields in the
convection zone and cyclic dynamo activity \citep{Kapyla_et_al_2010}.
Interestingly, large-scale magnetic fields and 
organized polarity reversals have also been achieved in spherical
geometries with helically forced flows, and involving neither global-scale
differential rotation nor tachoclines \citep{Mitra_et_al_2010}.

Here we explore convection and dynamo action in a rapidly rotating sun
which spins five times faster than the current solar rate,
calling this case~D5.  This dynamo builds
globally organized fields but also undergoes quasi-regular reversals of
magnetic polarity.  This is achieved in the stellar
convection zone itself; a tachocline is not included in our
simulation and thus can play no role. 
We describe briefly how our simulations are conducted in
Section~\ref{sec:ASH} and then compare cyclic case~D5 with the dynamo
case~D3 from \mbox{Paper I} in Section~\ref{sec:D3 and D5}.
In Section \ref{sec:D5 convection} we examine the nature of convection
in these rapidly rotating suns.  The magnetic fields and their
global-scale polarity reversals are explored in Sections
\ref{sec:energies}--\ref{sec:D5 production}.   
The differential rotation shows marked signatures of the reversals,
and these torsional oscillations are discussed in
Section~\ref{sec:torsional oscillations}.   Unusual magnetic structures arise
during certain intervals, and these are briefly discussed in
Section~\ref{sec:D5_strange_states}.   We reflect on our findings in
Section~\ref{sec:conclusions}.

\newpage
\section{Simulating Stellar Convective Dynamos}
\label{sec:ASH}

We study MHD stellar convection and dynamo action with the anelastic
spherical harmonic (ASH) code \citep{Clune_et_al_1999, Brun_et_al_2004}.
Our simulation approach is briefly described here, but the equations,
implementation and nature of our study are explained in detail in
\mbox{Paper I} \citep{Brown_et_al_2010a}. 
This global-scale code simulates a stratified spherical shell and here we focus
on the bulk of the convection zone, with our computational domain
extending from $0.72\:R_\odot$ to $0.97\:R_\odot$ (with $R_\odot$ the
solar radius) and spanning about
172~Mm in radius.  We avoid regions near the stellar surface and also
do not include the stably stratified region near the base of the
convection known as the tachocline.
The total density contrast across the shell is
about 25, with a stratification derived from a 1D solar structure
model \citep{Brown_et_al_2008}.  As discussed in \mbox{Paper I}
(Section~3.1), we chose to simplify these calculations by omitting a
tachocline of penetration and rotational shear, motivated also by the
lack of observational evidence for tachoclines in rapidly rotating
stars.  We have explored some companion simulations that include a
tachocline, and find that the wreath-building dynamics and their
temporal behavior are largely unmodified by its presence.  We will
report on this in another paper.  Furthermore, recent mean-field
models of the solar dynamo suggest that the latitudinal shear in the
lower convection zone is primarily responsible for the generation of
the mean toroidal field, as opposed to the radial shear in the
tachocline \citep{Dikpati&Gilman_2006, Rempel_2006,
Munoz-Jaramillo_et_al_2009}.  The role of a tachocline in establishing cyclic
magnetic activity remains unclear.

ASH is a large-eddy simulation (LES)
code with subgrid-scale (SGS) treatments for turbulent diffusivities.
As in \mbox{Paper I} the vorticity, entropy and magnetic field
diffusivities, $\nu$, $\kappa$ and $\eta$ respectively, are functions
of radius only and vary with the background density as
$\bar{\rho}^{-1/2}$.  Thus, all three diffusivities are smallest in
the lower convection zone. As in \mbox{Paper I}, the magnetic Prandtl
number $\mathrm{Pm}=\nu/\eta =0.5$.
The fundamental characteristics of our simulations and
parameter definitions are presented in Table~\ref{table:sim_parameters}.  

The dynamo simulation (case~D5) was initiated from a mature hydrodynamic
progenitor which had been evolved for more than 8000~days 
and was well equilibrated (case~H5).  
To initiate our dynamo case, a small seed dipole magnetic field was
introduced and evolved via the induction equation.  The energy in the
magnetic fields is initially many orders of magnitude smaller than the energy
contained in the convective motions, but these fields are amplified by
shear and grow to become comparable in energy to the convective
motions.  Our magnetic boundary conditions are a perfect conductor at
the bottom of the convection zone and match onto an external potential
field at the top.

Stellar dynamo simulations are computationally intensive, requiring both
high resolutions to correctly represent the velocity fields and long
time evolution to capture the equilibrated dynamo behavior, which may
include cyclic variations on time scales of several years.  
The strong magnetic fields can produce rapidly moving Alfv\'en waves
which seriously restrict the Courant-Friedrichs-Lewy (CFL)
timestep limits in the upper portions
of the convection zone.    Case~D5, rotating five times faster than the
current Sun, has been evolved for over 17000 days (or over 8 million
timesteps).  As a historical note, with the higher resolution of this
simulation, this represents roughly a factor of a million more
computation than was possible in \cite{Gilman_1983}, which is in
surprisingly good agreement with Moore's law doubling over the almost
thirty year interval separating these simulations.
We plan to report on a variety of other dynamo cases, some at higher
turbulence levels and rotation rates, in subsequent papers; cyclic
activity is a generic feature of many of these dynamos.

\begin{deluxetable*}{cccccccccccccc}
 \tabletypesize{\footnotesize}
  \tablecolumns{14}
  \tablewidth{0pt}  
  \tablecaption{Parameters for Primary Simulations
  \label{table:sim_parameters}}
  \tablehead{\colhead{Case}  &  
    \colhead{$N_r,N_\theta,N_\phi$} &
    \colhead{Ra} &
    \colhead{Ta} &
    \colhead{Re} &
    \colhead{Re$'$} &
    \colhead{Rm} &
    \colhead{Rm$'$} &
    \colhead{Ro} &
    \colhead{Ro$'$} &
    \colhead{Roc} &
    \colhead{$\nu$} &
    \colhead{$\eta$} &
    \colhead{$\Omega_0/\Omega_\sol$}
 }
 \startdata
  D5    & $96 \times 256 \times 512$ & 1.05$ \times 10^{  6}$ &     6.70$ \times 10^{  7}$ & 273 &  133 &  136 &   66 &    0.273 & 0.173 &  0.241 &    0.940 &     1.88 &  5 \\
  H5    & $96 \times 256 \times 512$ & 1.27$ \times 10^{  6}$ &     6.70$ \times 10^{  7}$ & 576 &  141 &  --- &  --- &    0.303 & 0.182 &  0.268 &    0.940 &     --- &  5 \\[3mm] 
  D3    & $96 \times 256 \times 512$ & 3.22$ \times 10^{  5}$ &     1.22$ \times 10^{  7}$ & 173 &  105 &   86 &   52 &    0.378 & 0.255 &  0.311 &     1.32 &     2.64 &  3 \\
  H3    & $96 \times 256 \times 512$ & 4.10$ \times 10^{  5}$ &     1.22$ \times 10^{  7}$ & 335 &  105 &  --- &  --- &    0.427 & 0.265 &  0.353 &     1.32 &     --- &  3 \\

   \enddata
\tablecomments{Dynamo simulation at 5 times the solar rotation
  rate is case D5, and its hydrodynamic (non-magnetic) companion is H5.
      Both simulations have inner radius 
	$r_\mathrm{bot} = 5.0 \times 10^{10}$cm and outer radius of 
      $r_\mathrm{top} = 6.72 \times 10^{10}$cm, with 
	$L = (r_\mathrm{top}-r_\mathrm{bot}) = 1.72 \times 10^{10}$cm
	the thickness of the spherical shell.
	Evaluated at mid-depth are the
	Rayleigh number $\mathrm{Ra} = (-\partial \rho / \partial S)
	(\mathrm{d}\bar{S}/\mathrm{d}r) g L^4/\rho \nu \kappa$, 
	the Taylor number $\mathrm{Ta} = 4 \Omega_0^2 L^4 / \nu^2$, 
	the rms Reynolds number $\mathrm{Re}  = v_\mathrm{rms} L /\nu$ and
	fluctuating Reynolds number $\mathrm{Re}' = v_\mathrm{rms}' L /\nu$,
	the magnetic Reynolds number $\mathrm{Rm} = v_\mathrm{rms} L/\eta$ 
	and fluctuating magnetic Reynolds number 
      $\mathrm{Rm}' = v_\mathrm{rms}' L/\eta$, 
	the Rossby number $\mathrm{Ro} = \omega / 2 \Omega_0$
	and fluctuating Rossby number $\mathrm{Ro}' = \omega' / 2 \Omega_0$,
	and the convective Rossby number 
	$\mathrm{Roc} = (\mathrm{Ra}/\mathrm{Ta} \, \mathrm{Pr})^{1/2}$.
	Here the fluctuating velocity $v'$ has the axisymmetric
        component removed: $v' = v - \langle v \rangle$, 
        with angle brackets denoting an average in longitude.  The
        rms velocities in the dynamo simulations (and corresponding Reynolds and Rossby numbers) are
        reduced because the differential rotation is weaker; the fluctuating
        velocities remain comparable.
	For both simulations, the Prandtl number $\mathrm{Pr} = \nu / \kappa$ is 0.25 
	and in the dynamo simulation the magnetic Prandtl number
      $\mathrm{Pm}=\nu/\eta$ is 0.5.   
	The viscous and magnetic diffusivity, $\nu$ and $\eta$, are
	quoted at mid-depth (in units of $10^{12}~\mathrm{cm}^2\mathrm{s}^{-1}$).
      The rotation rate $\Omega_0$ of each reference frame is in multiples
      of the solar rate $\Omega_\sol=2.6 \times
      10^{-6}~\mathrm{rad}\:\mathrm{s}^{-1}$ or $414$ nHz.   
      The~viscous time scale at mid-depth $\tau_\nu = L^2/\nu$ is
      about $3640$~days for case D5 and the resistive time scale is
      about $1820$~days, whereas the rotation period is 5.6~days.  For
      convenient reference, we repeat from \mbox{Paper I} the same
      data for cases D3 and H3 rotating at three times the solar rate.
	}
\end{deluxetable*}

\section{Dynamos in Rapidly Rotating Suns}
\label{sec:D3 and D5}

\begin{figure}
  \begin{center}
    \includegraphics{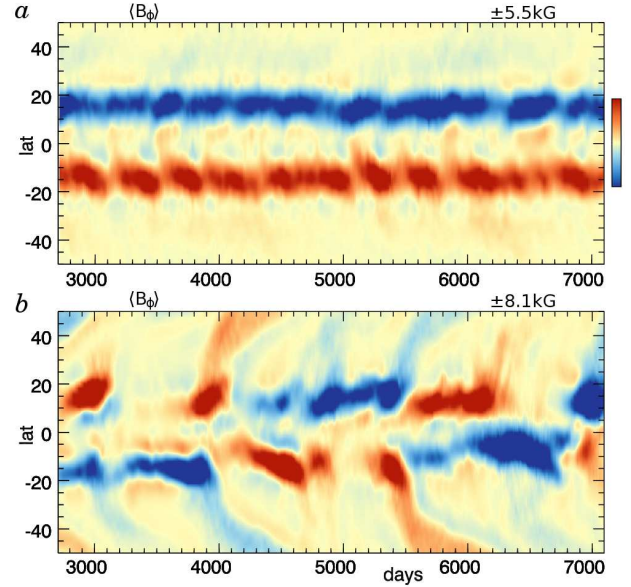}
  \end{center}
  \caption{Magnetic wreaths achieved in $(a)$~case~D3 (\mbox{Paper I})
    and $(b)$~case~D5 (this paper).  Shown are time-latitude plots of
    mean toroidal (longitudinal) magnetic field $\langle B_\phi \rangle$ at
    mid-convection zone, with scaling values indicated.  Case~D3
    builds persistent, time-independent wreaths but the wreaths
    achieved in case~D5 undergo quasi-regular reversals of polarity
    (three shown here, with roughly a 1500~day period).  The dynamic
    range of the color bars are indicated.
    \label{fig:D3 and D5 wreaths}} 
\end{figure}

In order to accentuate and investigate the 
self-organization processes associated with
helicity and rotational shear, we have conducted
a series of simulations of solar-type stars
rotating more rapidly than the Sun.  The 
non-magnetic analogues of our dynamo simulations
exhibit a systematic increase in the rotational
shear of differential rotation with increasing rotation rate.
Convection remains vigorous at 
all rotation rates, though at the highest rotation rates novel
localized nests of convection arise in the equatorial regions
\citep{Brown_et_al_2008}.

In \mbox{Paper I} \citep{Brown_et_al_2010a} we describe in detail 
the generation of persistent toroidal wreathes
in a simulation rotating at three times the 
solar rate (case~D3, at $3\: \Omega_\odot$).  These wreaths are localized
bands of strong ($\sim $ 7~kG, with peak amplitudes of $\sim$ 26~kG)
toroidal flux located in the midst of a turbulent convection zone,
sustained by rotational shear.  The toroidal field
strength peaks near the base of the convection zone
at latitudes of $\pm 15^\circ$, with opposite polarity 
in the northern and southern hemispheres.  
The mean toroidal (longitudinal) magnetic fields for case~D3 are shown in
Figure~\ref{fig:D3 and D5 wreaths}$a$. 
These magnetic structures form within 2000 days from weak initial seed
fields and persist for the remainder of the simulation,
spanning over 15,000 days of evolution.  They are nearly steady in
time and do not show global-scale reversals of magnetic polarity.

\newpage
Here we focus on another simulation of a solar-type
star with a faster rotation period of $5\: \Omega_\odot$ (case~D5).
As in Paper~I, the simulation builds strong wreaths
of toroidal field, but unlike those described
in Paper~I, these wreaths undergo quasi-periodic
polarity reversals.   Three such reversals are shown in 
Figure~\ref{fig:D3 and D5 wreaths}$b$.
During a cycle, the global-scale magnetic fields
wax and wane in strength and can flip their polarity.
These magnetic activity patterns
emerge spontaneously from a turbulent convective flow that 
is significantly more complex than in previous laminar,
Boussinesq simulations of cyclic dynamos.

\begin{figure*}
\begin{center}
  \includegraphics[width=\linewidth]{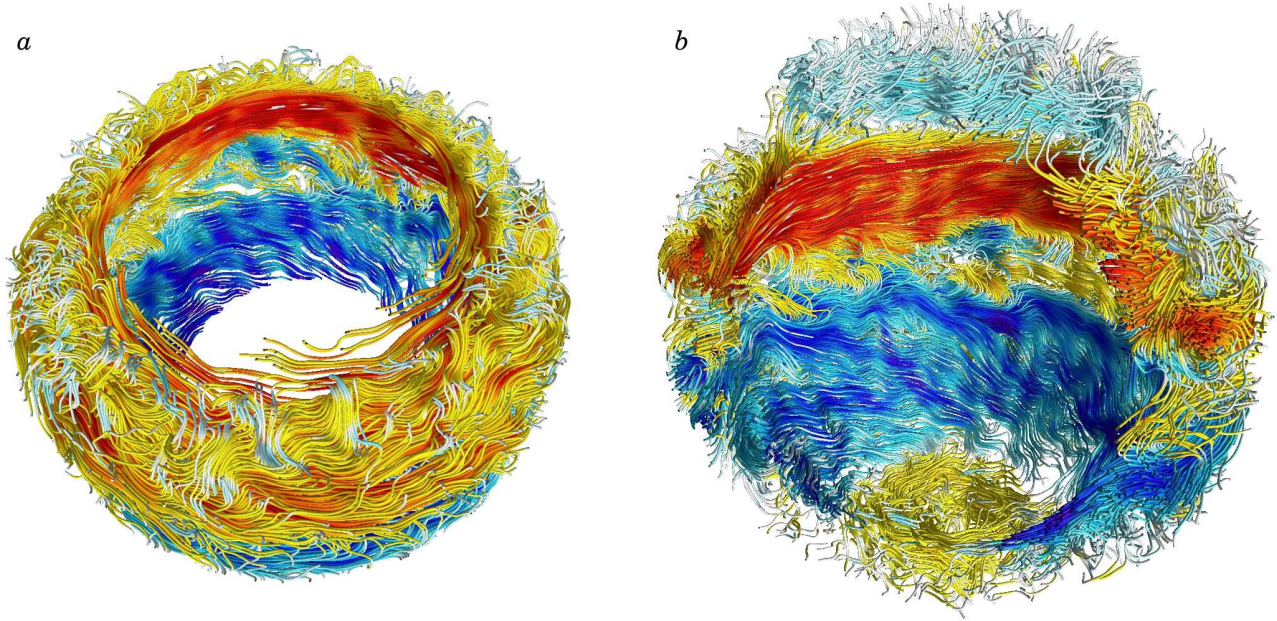}
\end{center}
\caption{Field line tracings of magnetic wreaths in case D5. $(a)$
  Snapshot of two wreaths in the equatorial region at day 3880
  (time~t1), when the magnetic fields are strong.  Lines trace the
  magnetic fields and color denotes the amplitude and polarity of
  the longitudinal field $B_\phi$ (red, positive; blue, negative);
  here the view is restricted to the equatorial region, spanning
  roughly $\pm 30^\circ$ latitude. Magnetic field threads in and out
  of two oppositely directed wreaths, with substantial connectivity
  across the equator.  $(b)$ Side view spanning slightly more than one
  hemisphere, showing connectivity from equatorial regions to polar
  caps.  Relic wreaths from the previous magnetic cycle are visible at
  the poles.
\label{fig:D5 wreaths}}
\end{figure*}

The wreaths of magnetism are highly
intricate structures with substantial connectivity throughout the
convection zone.  
The magnetic wreaths realized in case~D5 are shown in field line
tracings throughout the volume in Figure~\ref{fig:D5 wreaths} at a
time when the magnetic fields are strong.  
As in \mbox{Paper I}, we find that the wreaths are
topologically leaky structures, with magnetic fields threading in and
out of the main flux concentrations near the equator.  Rather than
being isolated entities, near the equator the two strong wreaths of
oppositely directed polarities have substantial cross-equatorial
connectivity (Fig.~\ref{fig:D5 wreaths}$a$).  This cross-equatorial
flux appears to play an important role in the polarity reversals that
are observed.  

Unlike the wreaths of case~D3, here magnetism fills the
entire convection zone including the polar regions 
(Fig.~\ref{fig:D5 wreaths}$b$). 
On their high-latitude (polar) edges, the wreaths near the equator are
connected to magnetic structures of weaker amplitude and opposite polarity at the
polar caps. These polar structures are relic wreaths from the previous
cycle that propagate towards the poles during the polarity reversal.
In a short time after this snapshot, the strong wreaths near the
equator begin to propagate towards the poles and are replaced by new
wreaths of opposite polarity (blue in northern and red in southern
hemisphere).  This phenomena is visible in Figure~\ref{fig:D3 and D5
  wreaths}$b$ starting at roughly day~4000, with the reversal
completed a short time later.  This is a remarkable example of
magnetic self-organization by turbulent, rotating, stratified convection that bears 
strongly on the vibrant magnetic activity and cyclic
variability observed in many young, rapidly-rotating
stars.

\section{Patterns of Convection in Case~D5}
\label{sec:D5 convection}

\begin{figure}
  \begin{center}
    \includegraphics[width=8cm]{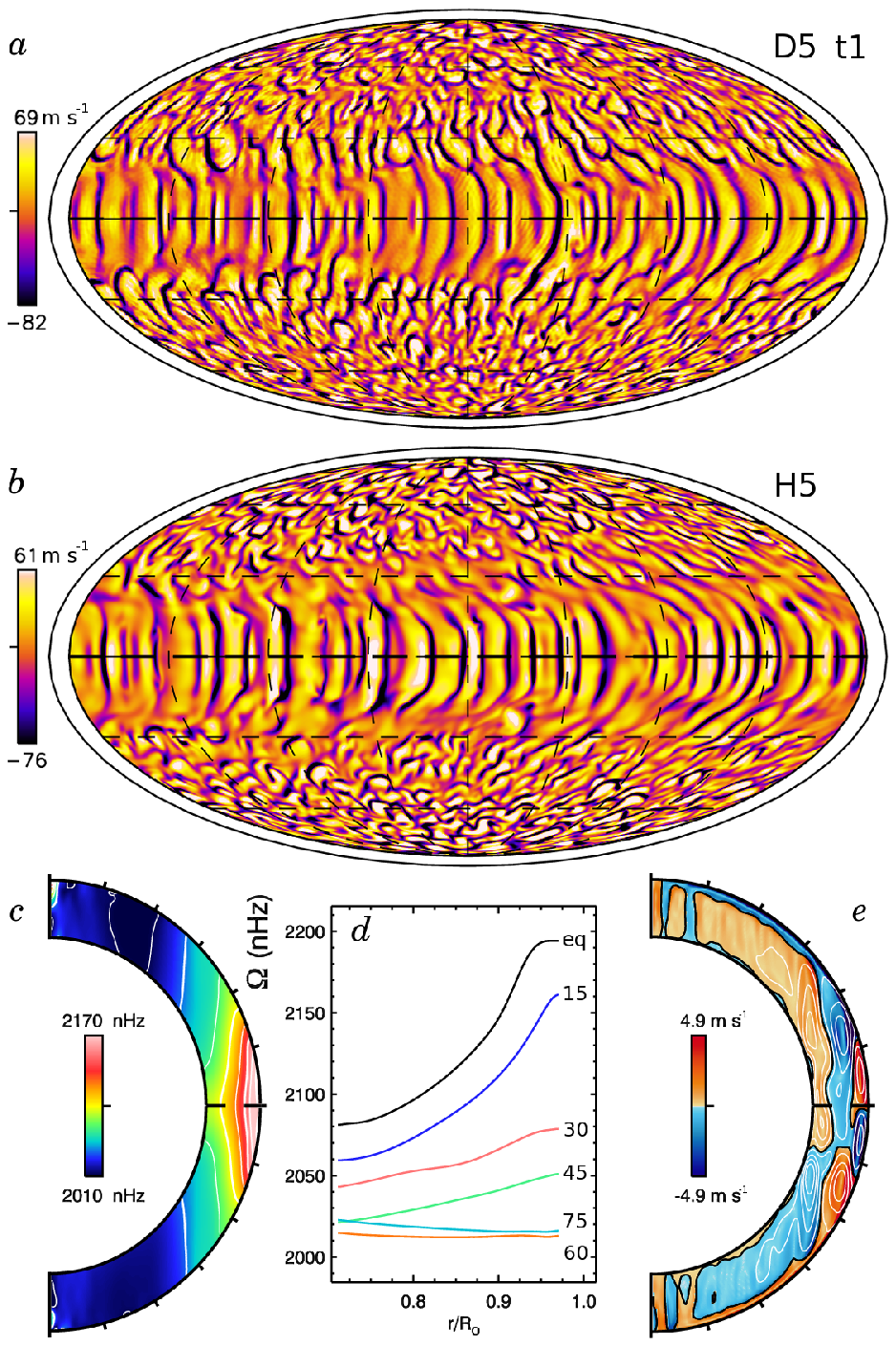}
  \end{center}
  \caption
     {Convective structures and mean flows in cases~D5 and H5.
     $(a)$~Radial velocity $v_r$ in dynamo case~D5 shown in
     global Mollweide projection at $0.95R_\odot$ with upflows light
     and downflows dark.  Poles are at top and bottom and the equator
     is the thick dashed line, while the stellar surface at $R_\odot$
     is indicated by the thin surrounding line.  This snapshot samples
     day 3880 (time~t1) when the magnetic fields are strong.
     $(b)$~Companion hydrodynamic
     case~H5.  Here stronger differential rotation shears out convective
     structures in the mid-latitudes.  $(c)$~Profile of mean angular
     velocity $\Omega(r,\theta)$ for case~D5, with $(d)$~radial cuts of
     $\Omega$ at selected latitudes.  $(e)$~Meridional circulations for
     case~D5, with magnitude and sense of circulation indicated by color
     (red counter-clockwise, blue clockwise) and streamlines of mass flux
     overlaid.  The profiles  shown in $(c-e)$ have been averaged over
     nearly 2000~days, spanning a full magnetic reversal.
  \label{fig:case_D5_patterns}}
\end{figure}

The complex patterns of convection of our dynamo and hydrodynamic 
cases rotating at five times the solar rate are presented in    
Figure~\ref{fig:case_D5_patterns}.
Individual convective cells are shown in snapshots of radial
velocity near the top of the simulated domain in
Figures~\ref{fig:case_D5_patterns}$a$  and $b$ for cases~D5 and H5
respectively.  Both cases share strong similarities in their
convective patterns.  Owing to the density stratification the
convection is compressible and the downflows are narrow and fast, while
the upflows are broader and slower.  

Near the equator the prominent cells are aligned
north-south and propagate in the prograde direction.  
The strongest flows span the entire convection zone;
the weaker cells are partially truncated by the strong zonal flows
of differential rotation.  Nearer to the poles (above roughly $\pm
45^\circ$ latitude) the patterns are more isotropic.
Networks of downflow lanes surround upflows and both propagate in a
retrograde fashion.  There is less radial shear and most of the convective
cells span the full convection zone.  
In the polar regions, the radial velocity patterns have a somewhat cuspy
appearance, with the strongest downflows appearing to favor the
westward and lower-latitude side of each convective cell.  This may be
a consequence of the strong retrograde differential rotation in those
regions.  

The convective downflow structures propagate
more rapidly than the differential rotation that they establish and in
which they are embedded. In the equatorial band, these structures move in
a prograde fashion and at high latitudes in a retrograde sense.
Individual convective cells typically persist for about 10 days,
though some have much longer lifetimes. 

The convective structures in case~D5 are quite similar to those realized in the
hydrodynamic case~H5 (Figure~\ref{fig:case_D5_patterns}$b$), though
there are some noticeable differences, particularly at the mid
latitudes (around $\pm30^\circ$). In the hydrodynamic case there is
little radial flow in these regions, as the strong differential
rotation shears out the convective cells.  This region is
equatorward of the tangent cylinder, an imaginary boundary tangent to
the base of the convection zone and aligned with the rotation axis.
For rotating convective shells, it has generally been found that the
dynamics are different inside and outside the tangent cylinder, due to
differences in connectivity and rotational constraint in these two
regions \citep[e.g.,][]{Busse_1970}.  The tangent cylinder in our geometry
intersects with the stellar surface at roughly $\pm42^\circ$ of
latitude.  In our compressible simulations, we generally find that the
convective patterns in the equatorial regions are bounded by a conic
surface rather than the tangent cylinder \citep{Brown_et_al_2008}.  In
case~H5 the strong differential rotation serves to disrupt the
convection at the mid-latitudes.  In contrast, in the dynamo case~D5
the differential rotation is substantially weaker in both radial and
latitudinal angular velocity contrasts.  As is evident in
Figure~\ref{fig:case_D5_patterns}$a$, the convective cells fill in
this region quite completely.

In our prior hydrodynamic simulations of convection in younger suns we
reported on localized nests of convection \citep{Brown_et_al_2008},
with those most prominent at the highest rotation rates.  Though there
is some modulation with 
longitude in the equatorial roll amplitudes here, this modulation is
less extreme in either case~D5 or H5 than in our previous rapidly
rotating simulations of stellar convection \citep[e.g., case~G5 in
][]{Brown_et_al_2008}.  This difference appears to be linked to our
background stratification and feedbacks from thermal transport near the
top of the domain.  Here we have attempted to
reduce the region of influence of the unresolved SGS heat flux $F_u$
which carries flux out the top of the domain \citep{Brown_et_al_2008}.
This thinner thermal boundary has larger gradients and a steeper
profile of the background entropy gradient
$\mathrm{d}\bar{S}/\mathrm{d}r$ and 
thus slightly higher Rayleigh numbers and radial velocities.  In our
broader study of rapidly rotating dynamos we have found that strongly
localized active nests of convection remain possible in dynamo
simulations at the most rapid rotation rates
 ($\Omega \gtrsim 10\:\Omega_\odot$).

The convection establishes a prominent solar-like differential
rotation, with a fast prograde equator and slow retrograde poles.
Figure~\ref{fig:case_D5_patterns}$c$ shows the profile of mean angular
velocity realized in case~D5, averaged in azimuth (longitude) and time
over a period of roughly 200 days centered on the time of the snapshot
in Figure~\ref{fig:case_D5_patterns}$a$.  The equatorial acceleration
is achieved by Reynolds stresses and convective transport that
redistribute angular momentum and entropy, and build prominent
gradients in latitude of angular velocity and temperature.  Radial
cuts of $\Omega$ indicate that strong radial shear is present
throughout the lower-latitudes (Figure~\ref{fig:case_D5_patterns}$d$).
This differential rotation is solar-like in the sense that there is a
monotonic decrease of $\Omega$ from the equator to the pole.
Generally, the profiles here of $\Omega$ are more cylindrical than
those deduced from helioseismology for the Sun, but this is to be
expected for more rapidly rotating stars. This may also be influenced
by our omission of a  tachocline.  From studies of solar convection,
it is evident that the thermal structure of the tachocline with
latitude influences the differential rotation profiles in the bulk of
the convection zone. The main effect is to tilt the $\Omega$ contours
toward a more radial alignment \citep{Rempel_2005, Miesch_et_al_2006}.

The differential rotation achieved is stronger in our hydrodynamic
case~H5 than in our dynamo case~D5.
This can be quantified by measurements of the 
latitudinal angular velocity shear $\Delta \Omega_\mathrm{lat}$.  Here, as in
\cite{Brown_et_al_2008} and \mbox{Paper I}, we define
$\Delta \Omega_\mathrm{lat}$ as the shear near 
the surface between the equator and a high latitude, 
say~\mbox{$\pm60^\circ$}, with
\begin{equation}
  \Delta \Omega_\mathrm{lat} = \Omega_\mathrm{eq} - \Omega_{60},
  \label{eq:absolute_contrast}
\end{equation}
and the radial shear $\Delta \Omega_\mathrm{r}$ as the angular
velocity shear between the surface and bottom of the convection zone
near the equator with 
\begin{equation}
  \Delta \Omega_\mathrm{r} = \Omega_{0.97R_\odot} - \Omega_{0.72R_\odot}.
  \label{eq:absolute_radial_contrast}
\end{equation}
We further define the relative shear as 
$\Delta \Omega_\mathrm{lat}/\Omega_\mathrm{eq}$.
In both definitions, we average the measurements of $\Delta \Omega$
in the northern and southern hemispheres, as the rotation profile is
often slightly asymmetric about the equator.

These measurements are quoted for case~D5 and H5 in
Table~\ref{table:delta_omega}.  
The angular velocity shears in case~D5 can vary substantially in
time as the magnetic fields wax and wane in strength.  As such, here
we quote measurements during one magnetic cycle with measurements
averaged over the entire time interval (with label \emph{avg} and date
range indicated) and at periods of strongest and weakest differential
rotation during this cycle (\emph{max} and \emph{min} respectively, at indicated times).
In the hydrodynamic simulation the differential rotation shows far
smaller time-variations.  The global-scale magnetic fields realized in
the dynamo case~D5 feed back on the differential rotation and strongly
diminish the amplitude of the angular velocity shears as compared
with the hydrodynamic case~H5. This results from both a slowing of the
equatorial rotation rate and an increase in the rotation rate in the polar regions.  

\newpage
The angular velocity shear realized in case~D5 in both latitude
$\Delta \Omega_\mathrm{lat}$ and radius $\Delta \Omega_r$ 
is remarkably similar in amplitude to that realized in our previous
dynamo simulation case~D3 (\mbox{Paper I}) even though the basic rotation rate
$\Omega_0$ is substantially faster.  This is in striking and marked contrast to our
hydrodynamic companion cases (H5 and H3) where faster rotation leads to greater
angular velocity contrasts (Table~\ref{table:delta_omega}).

\begin{deluxetable}{lcccc}
 \tabletypesize{\footnotesize}
  \tablecolumns{5}
  \tablewidth{0pt}  
  \tablecaption{Global Properties of Angular Velocity
  \label{table:delta_omega}}
\tablehead{\colhead{Case}  &  
    \colhead{$\Delta \Omega_\mathrm{lat}$} &
    \colhead{$\Delta \Omega_\mathrm{r}$} &
    \colhead{$\Delta \Omega_\mathrm{lat}/\Omega_\mathrm{eq}$} &
    \colhead{Epoch}
 }
 \startdata
  D5$^\text{{avg}}$     & 1.14 & 0.71 & 0.083 & 3500--5500 \\
  D5$^\text{{min}}$     & 0.91 & 0.39 & 0.067 & 3702 \\
  D5$^\text{{max}}$    & 1.43 & 0.98 & 0.102 & 4060 \\
  H5                           & 2.77 & 1.31 & 0.192 &  - \\[3mm]
   D3                       & 1.18 & 0.71 & 0.137 & 2010-6980 \\
   H3                       & 2.22 & 0.94 & 0.246 &  -  
  \enddata
  \tablecomments{Angular velocity shear in units of 
 $\mu \mathrm{rad}\: \mathrm{s}^{-1}$, with 
$\Delta \Omega_\mathrm{lat}$ measured near the surface 
 ($0.97R_\odot$) and $\Delta \Omega_\mathrm{r}$ measured across the
 full shell at the equator.  The relative latitudinal shear $\Delta
 \Omega_\mathrm{lat}/\Omega_\mathrm{eq}$ is also shown.
 For the dynamo cases, these measurements are taken over the indicated range of days.  In
 oscillating dynamo case~D5, these measurements are averaged over a long
 epoch ({avg}), and are also taken at two short intervals in time when the
 differential rotation is particularly strong ({max}) and when
 magnetic fields have suppressed this flow ({min}).   
The hydrodynamic case H5 is averaged for roughly 300 days and shows no
 systematic variation on longer timescales.  Measurements for cases
 D3 and H3 rotating at $3\:\Omega_\odot$ are quoted from \mbox{Paper I}.}
\end{deluxetable}

The differential rotation profiles realized in these rapidly rotating simulations are
substantially in thermal wind balance
\citep[e.g.,][]{Brun&Toomre_2002, Miesch_et_al_2006,
Brown_et_al_2008}, though large departures do arise near the
inner and outer boundaries where Reynolds stresses and boundary
conditions play a dominant role. Maxwell stresses are significant in
the cores of the magnetic wreaths realized in dynamo case~D5, but
these play relatively little role in the global transport of angular
momentum.  During a reversal, these Maxwell stresses
do however transport angular momentum toward the poles, giving rise
to bands of quickly flowing fluid that share some similarities with the
torsional oscillations observed during the solar cycle.  This behavior
will be explored in Section~\ref{sec:torsional oscillations}.

The meridional circulation patterns for case~D5  are shown in
Figure~\ref{fig:case_D5_patterns}$e$.  There appear to be three
major cells of circulation in each hemisphere, with several
cells in both radius and latitude.   Strong rotational constraint in
regions outside the tangent cylinder likely leads to the significantly
cylindrical nature of these slow flows.  Some flows along the inner
and outer boundaries cross the tangent cylinder and serve to weakly
couple the polar regions to the equatorial regions.  These meridional
circulations are very similar 
to the circulations found in case~H5. In both simulations, the
meridional circulations are weaker, slower and more multi-celled than
those realized in simulations rotating at the slower rotation rates
\citep{Brown_et_al_2008, Brown_et_al_2010a}.  The slower flows and
multi-cellular nature of these meridional circulations may have strong
implications for flux transport dynamos 
\citep[e.g.,][]{Bonanno_et_al_2006, Jouve&Brun_2007, Jouve_et_al_2010}.

\begin{figure*}
\begin{center}
  \includegraphics[width=\linewidth]{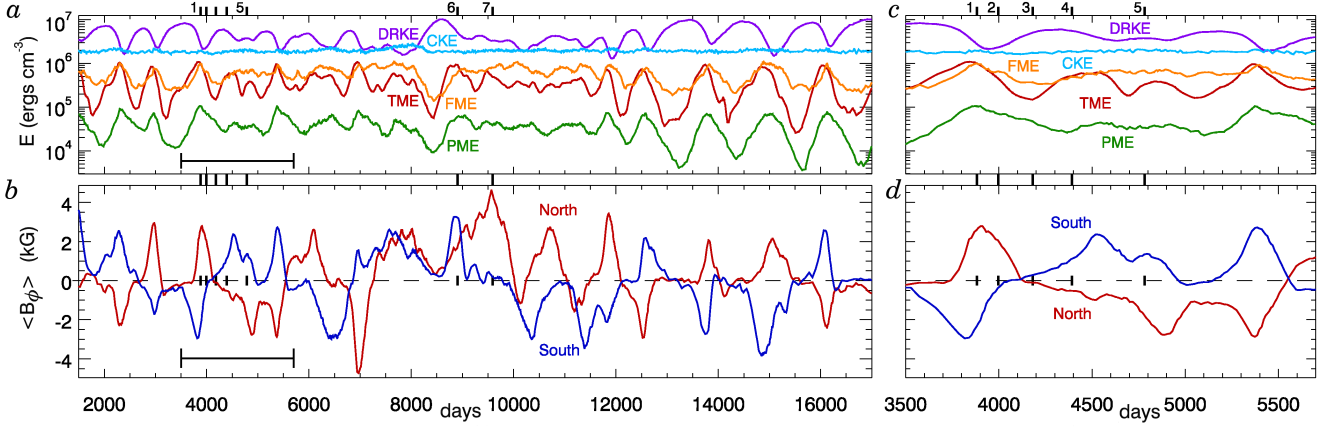}
\end{center}
\caption[Complex time evolution in case~D5 with flips in polarity of magnetic wreaths]
  {Complex time evolution in case~D5 with flips in polarity of magnetic wreaths. 
  $(a)$~Volume-averaged energy densities for kinetic
  energy of convection (CKE), differential rotation (DRKE) and for
  magnetic energy in fluctuating fields (FME), in mean toroidal fields
  (TME) and in mean poloidal fields (PME) as labeled.  Oscillations on
  roughly 500-1000 day periods are visible in the magnetic energies
  and in DRKE, though CKE stays nearly constant.  
  $(b)$~Mean toroidal field $\langle B_\phi \rangle$ averaged over entire northern and
  southern hemispheres (labeled) at mid-convection zone ($0.85 R_\sol$).
  Early in the simulation, opposite polarities dominate each
  hemisphere.  Several reversals occur, along with several extreme
  excursions which do not flip the polarity of the global-scale field.
  During the interval from roughly day 7700 to 10200, 
  the dynamo falls into peculiar single
  polarity states, with one polarity dominating both hemispheres.
  Bracketed interval from day 3500 to 5700 spans one full polarity
  reversal; $(c)$~expanded view of volume-averaged energy densities during
  this period, and $(d)$~the mean toroidal field with same vertical
  axis scales as in $(a,b)$.   Thick labeled tick marks above $a,c$
  indicate time samples used in later images. 
  %
\label{fig:case_D5_energies}} 
\end{figure*}

\newpage
\section{Oscillations in Energies and Changes of Polarity}
\label{sec:energies}

A striking feature of the convective dynamo case~D5 is its time
dependence.  This time-varying behavior is readily
visible as oscillations of the volume-averaged kinetic and magnetic
energy densities, as shown in Figure~\ref{fig:case_D5_energies}$a$ at
a time long after the dynamo has saturated and reached equilibration.

Here the kinetic energy of differential rotation (DRKE) undergoes
factor of five changes on periods of 500-1000 days.  As DRKE decreases
the magnetic energies increase. Moving in concert are the mean
toroidal (TME) and mean poloidal (PME) magnetic energies.  The mean
poloidal fields appear to lag slightly behind the mean toroidal fields
as they both change in strength. The fluctuating magnetic energy
(FME) tracks the largest rises in the mean fields but decouples during
many of the deepest dips.  In contrast, the variations in convective
kinetic energy (CKE) shows little organized behavior in time, and
appears to change substantially only when the differential rotation is
highly suppressed during the period from day 7500 to day 8300.  The
energy contained in the meridional circulations (MCKE) is weaker and
not shown.  Though it varies somewhat in time, there is not a clear
relation to the changes in magnetic energies.  These energies are defined as
\begin{eqnarray}
\mathrm{CKE}  &=& \frac{1}{2}\bar{\rho}\Big[\left(v_r - \langle v_r \rangle\right)^2+\left(v_\theta - \langle v_\theta \rangle\right)^2+\nonumber\\
              & & \qquad \left(v_\phi - \langle v_\phi \rangle\right)^2\Big], \\
\mathrm{DRKE} &=& \frac{1}{2}\bar{\rho}\langle v_\phi \rangle^2, \\
\mathrm{MCKE} &=& \frac{1}{2}\bar{\rho}\Big(\langle v_r \rangle^2 + \langle v_\theta \rangle^2 \Big),\\
\mathrm{FME}  &=& \frac{1}{8\pi}\Big[\left(B_r - \langle B_r \rangle\right)^2+\left(B_\theta - \langle B_\theta \rangle\right)^2+\nonumber\\
              & & \qquad \left(B_\phi - \langle B_\phi \rangle\right)^2\Big], \\
\mathrm{TME}  &=& \frac{1}{8\pi}\langle B_\phi \rangle^2, 
\end{eqnarray}
\begin{eqnarray}
\mathrm{PME}  &=& \frac{1}{8\pi}\Big(\langle B_r \rangle^2 + \langle B_\theta \rangle^2 \Big).
\end{eqnarray}
where angle brackets will consistently denote an average in longitude.

Magnetic energies in case~D5 can rise to be a substantial
fraction of the kinetic energies.  Averaged over the nearly 16000~days
(about 44~years) shown here, the magnetic energies are about 17\% of the kinetic
energies.  During individual oscillations the magnetic energies can
range from a few percent of the kinetic energies to levels as high as
50\%.  The kinetic energy is largely in the fluctuating convection and
differential rotation, with CKE fairly constant and ranging from
15-60\% of the total kinetic energy as DRKE grows and subsides, itself
contributing between 40 to 85\% of the kinetic energy.  The magnetic
energies are largely split between the mean toroidal fields and the
fluctuating fields, with TME containing about 35\% of the magnetic
energy on average, FME containing about 61\% and PME containing 4\%.
The roles of these energy reservoirs change somewhat through each
oscillation.  At any one time, between 10 and 60\% of the magnetic
energy is in TME while FME contains between 30 and 85\% of the total.
Meanwhile, PME can comprise as little as 1\% or as much as 10\% of the
total.  Generally, PME is about 12\% of TME, but because PME lags the
changes in TME slightly, there are periods of time when PME is almost
40\% of TME.

These results are in contrast to our previous simulations of the solar
dynamo, where the mean fields contained only a small fraction of the
magnetic energy \citep[e.g.,][ where TME and PME comprise about 2\% of the total]{Brun_et_al_2004}.
Simulations of dynamo action in fully-convective
M-stars do however show high levels of magnetic energy in the mean
fields \citep{Browning_2008}. In those simulations the fluctuating
fields still contain much of the magnetic energy,
but the mean toroidal fields possess about 18\% of the total
throughout most of the stellar interior.
Simulations of dynamo action in the convecting cores of A-type stars
\citep{Brun_et_al_2005} achieve similar results though when fossil
fields are included, those dynamos can reach states with
super-equipartition magnetic field strengths and strong mean fields
\citep{Featherstone_et_al_2009}. In our rapidly rotating suns, the mean fields
comprise a significant portion of the magnetic energy in the
convection zone and are as important as the fluctuating fields.
In the wreath-building dynamo case~D3 rotating at three times the current solar rate
(Paper~I) we found that the mean toroidal fields contained roughly 43\% of the
magnetic energy and the mean poloidal fields contained about 4\%, 
with the rest in the fluctuating fields.

The global-scale magnetic fields can reverse their polarities during
some of the oscillations in magnetic energies.  This is evident in
Figure~\ref{fig:case_D5_energies}$b$ showing
averages at mid-convection zone of the longitudinal magnetic
field $\langle B_\phi \rangle$ over the northern and southern
hemispheres.  Reversals  in field polarity occur periodically, with
typical time scales of roughly 1500 days.  These reversals appear to
happen shortly after peak 
magnetic energies are achieved, but do not occur every time magnetic
energies undergo a full oscillation.  Rather,  for each successful
polarity reversal it appears that several
failed reversals occur where the magnetic energies drop and the
average fields decline in strength, only to return with the same
polarity a few hundred days later.

We focus in the following discussion on one such reversal, shown in
closeup in Figures~\ref{fig:case_D5_energies}$c,d$ and spanning the
interval of time between days 3500 and 5700.  Two reversals occur during
this interval, with the global polarities flipping into a
new state at roughly day 4100 and then changing back again at about
day 5500.  Detailed measurements of kinetic and magnetic energies
during this interval are shown in Table~\ref{table:energies}. 

\begin{deluxetable}{lccccccc}
   \tabletypesize{\footnotesize}
    \tablecolumns{7}
    \tablewidth{0pt}  
    \tablecaption{Kinetic and Magnetic Energies
    \label{table:energies}}
    \tablehead{\colhead{Case}  &  
      \colhead{CKE} &
      \colhead{DRKE} &
      \colhead{MCKE} &
      \colhead{FME} &
      \colhead{TME} &
      \colhead{PME} 
   }
   \startdata
    D5$^\text{{avg}}$    & $1.85$ & $4.46$ & $0.006$ &
                               $0.55$ & $0.43$ & $0.048$ \\ 

    D5$^\text{{min}}$    & $1.70$ & $2.85$ & $0.005$ &
                               $0.50$ & $0.25$ & $0.062$ \\ 

    D5$^\text{{max}}$    & $1.85$ & $7.52$ & $0.007$ &
                               $0.39$ & $0.65$ & $0.042$ \\ 

    H5                       & $2.27$ & $34.3\phn\phn$ & $0.008$ &
                                -     &   -    & - 
    \enddata
 \tablecomments{Volume-averaged energy densities relative to the
    rotating coordinate system.  Kinetic energies are shown for
    convection (CKE), differential rotation (DRKE) and meridional
    circulations (MCKE).  Magnetic energies are shown for fluctuating
    magnetic fields (FME), mean toroidal fields (TME) and mean
    poloidal fields (PME).  All energy densities are reported in units of
    $10^{6} \mathrm{erg}\:\mathrm{cm}^{-3}$.  In time-varying case~D5,
    these energies are averaged over the intervals defined 
    in Table~\ref{table:delta_omega}, while in case~H5 an arbitrary
    1000~day interval was chosen.}
\end{deluxetable}

\section{Global-Scale Magnetic Reversals}
\label{sec:toroidal reversals}

\subsection{Toroidal Field Reversals}
\begin{figure}
  \begin{center}
  \includegraphics[width=\linewidth]{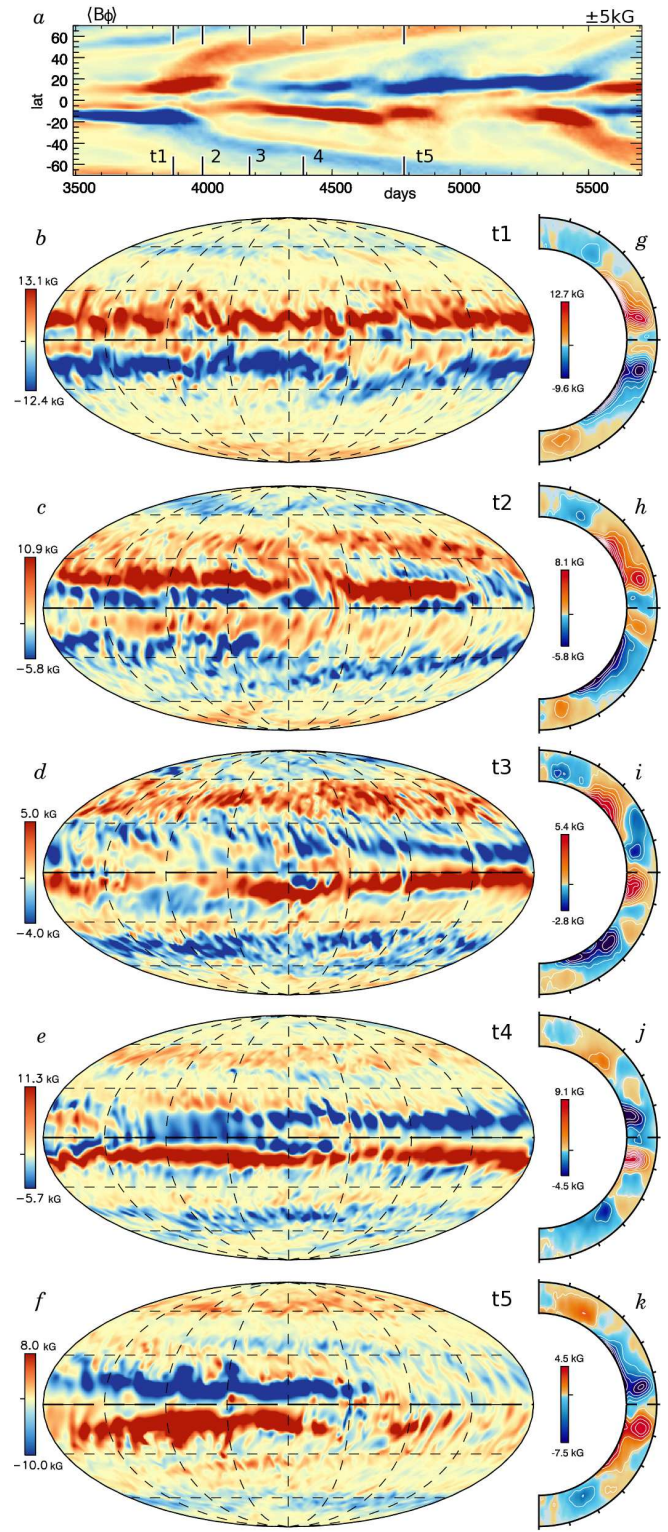}
  \end{center}
  \caption[Evolution of $B_\phi$ during a polarity reversal in case~D5]
          {Evolution of longitudinal field $B_\phi$ during a polarity reversal in case~D5.
    $(a)$~Time-latitude plot of $\langle B_\phi \rangle$ at
    mid-convection zone, with sampling times indicated.  
    $(b-f)$~Snapshots of $B_\phi$ in Mollweide
    projection at mid-convection zone ($0.85R_\odot$) at times
    indicated.  Between reversals the field is dominated by the
    mean component, but during reversals substantial fluctuations
    develop.  $(g-k)$~Accompanying samples of azimuthally-averaged 
    $\langle B_\phi \rangle$, showing structure of mean fields
    with radius and latitude at same instants in time.
    The companion poloidal fields are shown in Figure~\ref{fig:A_phi}.
  \label{fig:case_D5_reversal}}
\end{figure}

The nature of the global-scale magnetic fields during the reversal
spanning days 3500--5700
are presented in detail in Figure~\ref{fig:case_D5_reversal}.  Several
samples of longitudinal magnetic field $B_\phi$ are shown at mid-convection zone spanning this
time period.  The timing of these samples is indicated in
Figure~\ref{fig:case_D5_energies} by numeric labels and likewise in
Figure~\ref{fig:case_D5_reversal}$a$ which shows azimuthally-averaged
$\langle B_\phi \rangle$ at this depth in a time-latitude map that
spans the reversal.  The evolution of the mean toroidal field near the
base of the convection zone is discussed in Section~\ref{sec:D5 production} and Figure~\ref{fig:D5 production}$a$.

Before a reversal, the magnetic wreaths of case~D5 are very similar
in appearance to the wreaths realized in our persistent
wreath-building dynamo case~D3 (Paper~I).  They are dominated
by the azimuthally-averaged component of 
$B_\phi$, while also showing small-scale variations where convective
plumes distort the fields (Figure~\ref{fig:case_D5_reversal}$b$).
At mid-convection zone, typical longitudinal field strengths are of
order $\pm 13$~kG, while peak field strengths there can reach $\pm 40$~kG. 
Meanwhile $\langle B_\phi \rangle$ is fairly antisymmetric 
between the northern and southern hemispheres
(Figure~\ref{fig:case_D5_reversal}$g$).  Shortly before a reversal, the
magnetic wreaths strengthen in amplitude and become more
antisymmetric about the equator.

They reach their peak values just before the polarity change at
roughly day 4000 but then quickly begin to unravel, gaining
significant structure on smaller 
scales (Figure~\ref{fig:case_D5_reversal}$c$).  At the same time,
prominent magnetic structures detach from the higher-latitude edges and
begin migrating toward the polar regions.  
Meanwhile, $\langle B_\phi \rangle$  loses its antisymmetry between
the two hemispheres, with $\langle B_\phi \rangle$ in one hemisphere
typically remaining stronger and more concentrated than in the other
(Figure~\ref{fig:case_D5_reversal}$h$).  The stronger
hemisphere (here the northern) retains its polarity for about 100 days
as the fields in the other hemisphere (here southern) weaken and
reverse in polarity.  At this point the new wreaths of the next
cycle, with opposite polarity, are already faintly visible at the
equator. 

Within another 100 days these new wreaths grow in strength and become
comparable with the structures they replace, which are still visible at
higher latitudes (Figures~\ref{fig:case_D5_reversal}$d, i$).  The
mean $\langle B_\phi \rangle$ begins to contribute
significantly to the overall structure of the new wreaths, and soon
the polarity reversal is completed.  In the interval immediately after
the reversal, small-scale fluctuations still contribute significantly
to the overall structure of the wreaths, and $B_\phi$ has a complicated
structure at mid-convection zone.  At this time, the peak magnetic
field strengths are somewhat lower, at about $\pm 20$~kG. 
As $\langle B_\phi \rangle$ becomes stronger, the wreaths return to an
antisymmetric state, with 
similar amplitudes and structure in both the northern and southern
hemispheres (Figures~\ref{fig:case_D5_reversal}$e,j$).  They look much
as they did before the reversal, though now with opposite polarities.
  
The wreaths from the previous cycle appear to move through the lower
convection zone and toward higher latitudes.  This can be
seen variously in the time-latitude map at mid-convection zone
(Figure~\ref{fig:case_D5_reversal}$a$), in the Mollweide snapshots
(Figures~\ref{fig:case_D5_reversal}$b$--$f$), as well as in the 
samples of $\langle B_\phi \rangle$ (Figures~\ref{fig:case_D5_reversal}$g$--$k$).  
This poleward migration appears to be partially due to a dynamo wave
(Section~\ref{sec:D5 production})
and partially due to hoop stresses within the magnetic wreaths and an
associated poleward-slip instability 
\citep[e.g.,][and our Section~\ref{sec:torsional oscillations}]
{Spruit&vanBallegooijen_1982, Moreno-Insertis_et_al_1992}.
Even at late times some signatures of the previous wreaths persist in
the polar regions, and are still visible in Figures~\ref{fig:case_D5_reversal}$e,j$ at
day 4390.  They are much weaker in amplitude than the wreaths
at the equator, but they persist until the wreaths from the next
cycle move poleward and  replace them.   As they approach the polar
regions, the old wreaths dissipate on both large and small scales, for
the vortical polar convection shreds them apart and ohmic diffusion
reconnects them with the relic wreaths of the previous cycle.  The
$\Omega$-effect also contributes both to the low-latitude generation
of the wreaths and their high-latitude decay (Section~\ref{sec:D5 production}).

Though reversals occur on average once every 1500 days,
substantial variations can occur on shorter time scales.  
Here at mid-cycle the wreaths become concentrated in smaller
longitudinal intervals of the equatorial region (as in
Figures~\ref{fig:case_D5_reversal}$f,k$ at day 4780).  At other times 
the mean longitudinal fields become quite asymmetric, with one
hemisphere strong and one weak (i.e.,~during days 4900--5200) before
regaining their antisymmetric nature shortly prior to the next reversal.

\subsection{Connections Across the Equator}\label{sec:xeq}

\begin{figure}
  \begin{center}
  \includegraphics[width=\linewidth]{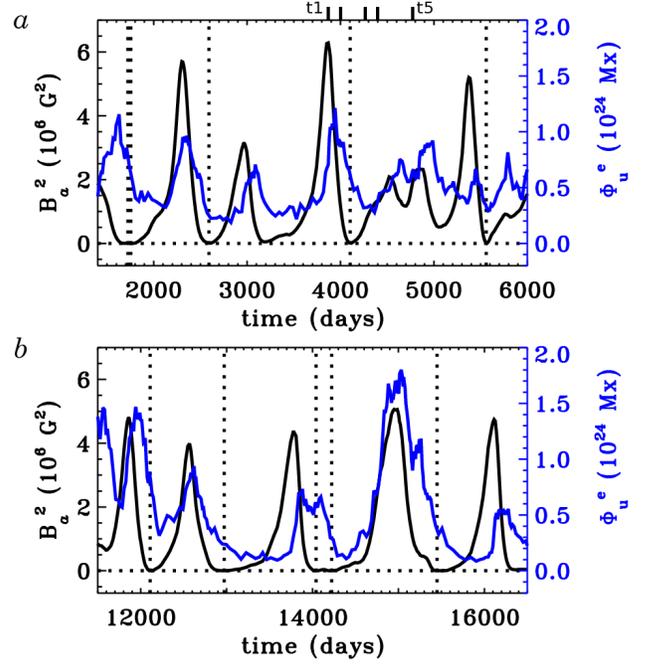}
 \end{center}
\caption{\label{fig:cross-equator} 
 The squared magnitude of the antisymmetric,
 low-latitude toroidal field component $B_a^2(t)$ (black lines,
 left axes) and the unsigned flux across the equator $\Phi_u^e(t)$
 (blue lines, right axes) as a function
 of time for two selected intervals when the wreaths of opposite
 polarity occupy each hemisphere: ($a$)
 $t = $ 1400--6000 (days), and ($b$) 11,500--16,500 (days). 
 Vertical dotted lines indicate polarity reversals, defined
 as where $B_a(t)$ changes sign. Labeled tick marks at top
 indicate times t1--t5.  Generally, the unsigned flux $\Phi_u^e(t)$ 
 is larger when $B_a^2$ is decreasing. During the interval
 6000--11,500 days (not shown) the
 dynamo is in an unusual state in which wreaths in both hemispheres have
 the same magnetic polarity. 
}
\end{figure}

The oppositely-directed wreaths are not isolated entities.
Rather, they interact through complex magnetic linkages
across the equator that contribute to their erosion and
subsequent reversal.

\begin{figure*}[t]
\begin{center}
  \includegraphics[width=\linewidth]{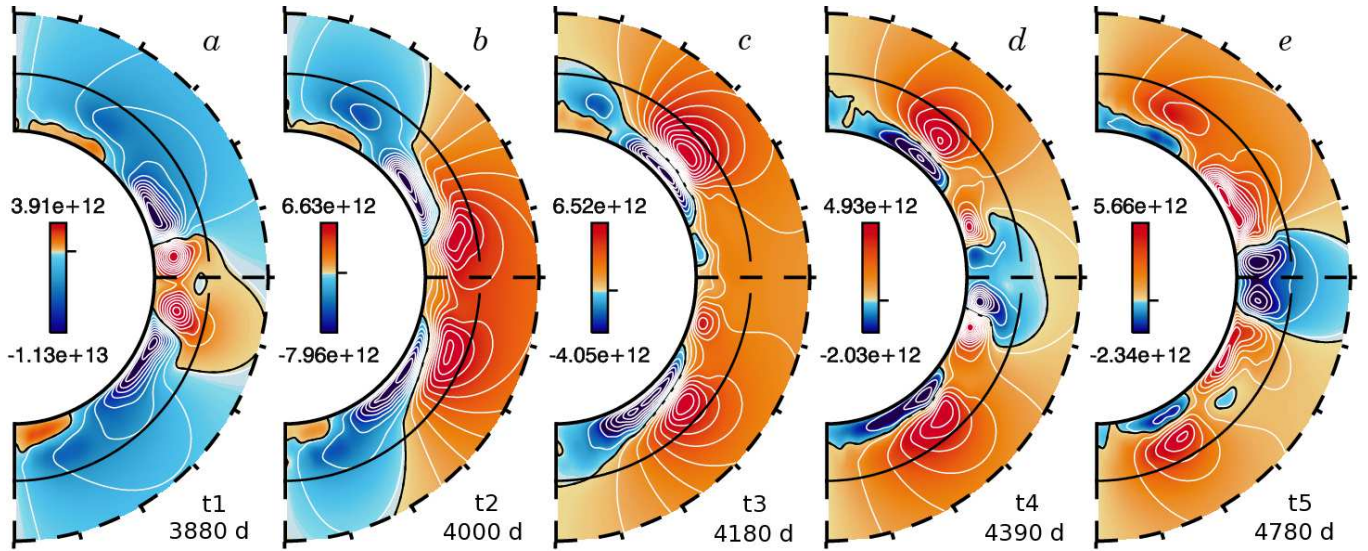}
\end{center}
\caption{Evolution of poloidal field during a polarity reversal in
 case~D5.  Shown are the vector potential of the mean poloidal field
 $\langle A_\phi \rangle$  in case D5 at times t1--t5 (with
 corresponding days noted).  Colors denote amplitude and polarity (red
 clockwise; blue, counterclockwise); units are G~cm.  The stellar
 surface is indicated with a solid line, and the poloidal field is
 extended with a potential field extrapolation to 1.3\:R$_\odot$
 (dashed surface).  Reversals of poloidal field appear to begin in the
 equatorial region $(a)$ and spread through the upper convection zone
 $(b,c)$ before new reversed polarity appears at the equator
 $(d,e)$.  The toroidal fields have largely reversed in the
 equatorial region by time~t3 (see~Figure~\ref{fig:case_D5_reversal}$d,i$).
 \label{fig:A_phi}}
\end{figure*}

The magnetic linkages between hemispheres can be quantified
by the net unsigned magnetic flux through the equatorial
plane:
\begin{equation}\label{uflux}
\Phi_u^e(t) = \int_{r_1}^{r_2} \int_0^{2\pi}
\left\vert B_\theta(r,\theta_{\mathrm{eq}},\phi) \right\vert r \sin\theta_e \mathrm{d}r \mathrm{d}\phi,
\end{equation}
where $\theta_{\mathrm{eq}} = \pi/2$ and where $r_1=r_{\mathrm{bot}}$ and
$r_2=r_\mathrm{top}$. 
In order to relate this quantity to polarity
reversals of the toroidal wreaths we define the antisymmetric component
of the mean, low-latitude toroidal field as follows
\begin{equation}\label{Banti}
B_a(t) = V^{-1} \int_{r_1}^{r_2} \int_{\theta_{\mathrm{eq}}-\delta}^{\theta_{\mathrm{eq}}+\delta}
\int_0^{2\pi} h(\theta) B_\phi(r,\theta,\phi) r^2 \sin \theta \mathrm{d}r \mathrm{d}\theta \mathrm{d}\phi \:,
\end{equation}
where
\begin{equation}
V = \int_{r_1}^{r_2} \int_{\theta_{\mathrm{eq}}-\delta}^{\theta_{\mathrm{eq}}+\delta} \int_0^{2\pi}
r^2 \sin\theta \mathrm{d}r \mathrm{d}\theta \mathrm{d}\phi = 
\frac{4 \pi}{3} \left(r_2^3 - r_1^3\right) \cos(\theta_e - \delta) \:.
\end{equation}
Here $h(\theta)$ is a step function with $h(\theta) = 1$ for
$\theta < \theta_e$, $h(\theta) = -1$ for
$\theta > \theta_e$, and $h(\theta_e) = 0$. We set $\delta = 2\pi/9$
(40$^\circ$) in order to focus on the low-latitude wreaths and then
plot the squared amplitude of this quantity.

Figure~\ref{fig:cross-equator} shows $\Phi_u^e(t)$ and $B_a^2(t)$ versus time for the
two intervals in the simulation when cyclic magnetic activity is
most apparent.  The interval spanning days 6000--11,500 is not shown as
the dynamo had fallen into a peculiar single polarity state and had
temporarily stopped showing cyclic behavior; we defer discussion of
that interval until Section~\ref{sec:D5_strange_states}. 
Labeled tick marks at the top of Figure~\ref{fig:cross-equator}$a$
indicate times t1--t5 shown 
previously during the reversal in Figure~\ref{fig:case_D5_reversal}
(days 3880 to 4780).

Examining this interval, we see that the squared amplitude of
antisymmetric mean field $B_a^2$ attains a 
peak value near time~t1 shortly before the reversal and then
drops to a minimum as the global-scale fields reverse in polarity
(reversal denoted by vertical dotted line at roughly day 4100).  
Some of this decrease is due to the poleward propagation of the
magnetic wreaths.  In Figure~\ref{fig:case_D5_reversal}$a$ we see that
at mid-convection zone, the wreaths of the
preceding cycle have largely left the region $\pm40^\circ$ latitude
by day~4100, leaving the wreaths of new polarity at the equator.
$B_a^2$ then slowly grows in amplitude before again attaining a sharp
maximum and reversing in sense (near day~5500).
The unsigned flux through the equator $\Phi_u^e$ lags
somewhat behind the mean fields, peaking near time~t2 and dropping to
a minimum at approximately day~4200.  The unsigned flux, measuring
cross-equatorial connectivity, does not drop to zero during any of the
intervals studied here. 

Clearly, the decay and subsequent reversal of
antisymmetric toroidal wreaths is strongly correlated with enhanced
magnetic linkages across the equator.  To quantify this relationship,
we define the rising and decaying phases of a cycle as those intervals
when the temporal derivative of $B_a^2(t)$ is positive and negative
respectively. Then we proceed to compute the average value of
$\Phi_u^e(t)$ in declining phases relative to rising phases. For the
two time intervals shown in Figure \ref{fig:cross-equator}, this ratio is ($a$) 2.7
and ($b$) 1.9.  Over the entire extended simulation interval (1400--18,252 days)
the ratio of unsigned flux in declining versus rising phases is 1.7.
Thus, there is significantly more unsigned flux across the equator
when the wreaths are declining in amplitude or undergoing a reversal.

\subsection{Poloidal Field Reversals}
\label{sec:poloidal reversals}
The evolution of the mean poloidal field can be followed by examining
how its vector potential $\langle A_\phi \rangle$ evolves in time, where
\begin{equation}\label{Bpol}
  \langle \vec{B}_\mathrm{pol} \rangle = \langle B_r \rangle \vec{\hat{r}} +
  \langle B_\theta \rangle  \vec{\hat{\theta}} = 
  \del \times \langle A_\phi \vec{\hat{\phi}} \rangle,
\end{equation}
with unit vectors denoted by hats and angle brackets again denoting an
average in longitude.
The vector potential of the mean poloidal magnetic field 
$\langle A_\phi \rangle$ is shown in snapshots at times t1--t5 during
the reversal in Figure~\ref{fig:A_phi}, with color 
denoting polarity and poloidal field lines represented by the
overlying contours.  A potential field extrapolation has been used to
follow the poloidal field above the surface out to a distance of
$1.3\:R_\odot$.  

When the magnetic fields are strong (i.e., at
time t1, Fig.~\ref{fig:A_phi}$a$) the poloidal field is dominated by
odd-$\ell$ components, with significant dipolar and octupolar
contributions.  The polar regions have the same polarity (here
negative), while the equatorial region has opposite polarity (here
positive).  The transition between poloidal polarities occurs in the
cores of the magnetic wreaths, where $\langle B_\phi \rangle$ is
strong (near~$\pm 15^\circ$ latitude).  

During the course of a reversal, $\langle A_\phi \rangle$ grows in
amplitude in the equatorial region.  The equatorial polarity expands
to fill the upper convection zone (times t2--t3,
Figs.~\ref{fig:A_phi}$b,c$) while in the lower convection zone the 
dominant polar polarity begins to disappear.  At the same times,
toroidal field of the new opposite polarity is appearing in this
equatorial region (Fig.~\ref{fig:case_D5_reversal}$c,d,h,i$).  Indeed,
by time~t3 the wreaths of the new cycle are well established and
the wreaths of the previous cycles are propagating towards the polar
regions.

The poloidal field of equatorial polarity from the previous cycle 
(positive in the reversal shown in Figure~\ref{fig:A_phi})
replaces the poloidal field at the poles, while in the equatorial
region oppositely directed poloidal field begins to appear at the
equator (here negative and visible at time~t4).   This occurs as the
new wreaths of toroidal field grow in strength and attain a high
degree of axial symmetry with $\langle B_\phi \rangle $ contributing
substantially to the structure of the toroidal field.
At this point the reversal is complete, though flux
continues to cancel in the lower convection zone, leading to the
final reversed state shown in Figure~\ref{fig:A_phi}$e$.

\section{Evolution of Large-Scale Moments}
\label{sec:global-scale reversals}  

The turbulent convection in case D5 gives rise both to small-scale, rapidly
evolving magnetism, and to slowly evolving ordered fields on larger scales.
One manifestation of the latter is the generation of wreaths of toroidal
field, whose strength and temporal evolution we have already described.
Another is the presence of remarkably strong dipole, quadrupole and octupolar
components of the poloidal field.  These low-order moments of the field do
not dominate the magnetic energy -- indeed, near the top of the simulation domain,
modes with spherical harmonic degree $\ell=1$--$3$ typically contain
no more energy than modes with $\ell$ up to $20$.  But these low-order
modes hold particular significance both theoretically and observationally: in the Sun, for
instance, the evolution of the global dipole moment is tightly linked to
the sunspot cycle, with the phasing relationship between the two (the sign
of the surface dipole reverses near solar maximum) serving as an important
constraint on models of the solar dynamo \citep[e.g.,][]{Wang&Sheeley_1991,
Charbonneau&MacGregor_1997}.  

More generally, the low-order modes of the magnetic field are
important both as diagnostics of global-scale dynamo action and as
mediators that regulate the interaction of a star with its
surroundings.  Because higher-order multipoles fall off quickly with
radius, the dipole mode dominates the magnetic energy distribution at
large distances from the stellar surface, and so may contribute most
to the magnetic ``lever arm'' that determines how rapidly a star's
rotation is braked
\citep[e.g.,][]{Weber&Davis_1967,Matt&Pudritz_2008}.  Low-order
multipoles also regulate the interaction of pre-main-sequence stars
with their surrounding accretion disks
\citep[e.g.,][]{Shu_et_al_1994}, with broad implications for the
truncation of those disks and the formation of protoplanetesimals.
Such considerations warrant a careful consideration of the 
multipolar structure and evolution in case D5.

\subsection{The Wandering Dipole}
\begin{figure}[t]
  \includegraphics[width=\linewidth]{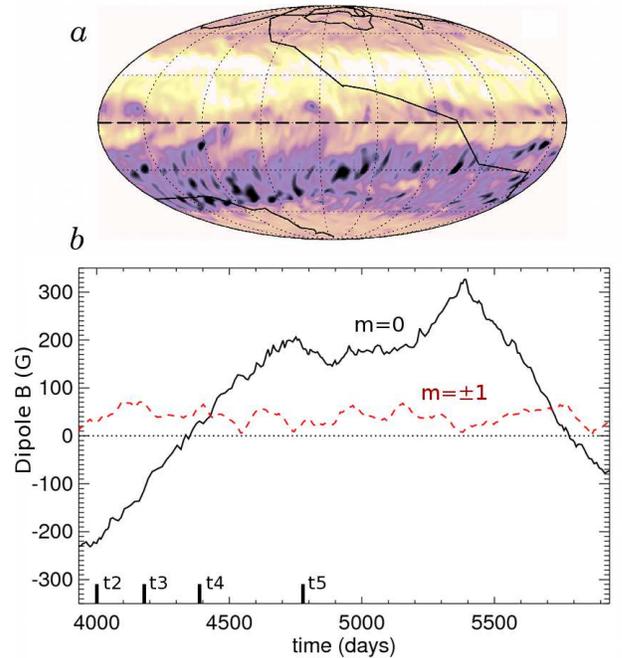}
\caption{
Evolution of dipole field in the lower-convection zone during one
magnetic reversal, with times t2--t5 noted.  
($a$) Location of the maximum in the dipole ($\ell=1$) magnetic field (the positive magnetic pole)
as measured at $0.73 R_\odot$ is traced by the solid black line, with $B_{\phi}$ at time~t2
and at the same depth shown in background.  ($b$) Relative strengths of the
axisymmetric ($m=0$) and non-axisymmetric ($m=\pm 1$, added in quadrature) terms of the
dipole.  The axisymmetric dipole smoothly evolves from negative to
positive during the reversal while the non-axisymmetric fields
fluctuate continuously and show little sign of the global-scale
polarity reversal.  A second reversal occurs after day 5500.
\label{fig:pole wander}}
\end{figure}

The lowest-order moment is the dipole, the components of which 
sum vectorally to give a time-evolving quantity with both magnitude
and direction.  Figure~\ref{fig:pole wander} illustrates the
the evolution of the dipole vector over the course of one 
polarity reversal.

The location of the positive magnetic pole in the lower-convection
zone is traced in Figure~\ref{fig:pole wander}$a$ over a period of
roughly 2000 days, displayed on a surface near the base of the 
convection zone. This track was calculated by filtering the radial field
$B_r$ in spectral space (retaining only $\ell=1$ modes), transforming
back to physical space, and tracking the longitude and latitude of the
maximum in the resulting field distribution.  In Figure~\ref{fig:pole
 wander}$a$, the latitude of the pole generally stays 
close to the geographic north or south pole, except during reversals of the
overall polarity.  These reversals are the only time during the interval
studied here that the pole ``tips'' to latitudes of less then about
$\pm 45$ degrees.  The longitudinal position of the pole fluctuates more
erratically, with no orderly sense of propagation.  
In examining both the longitude and latitude
of the $\ell=1$ maxima and thus including contributions from modes with
$m= \pm1$ as well as $m=0$, we are effectively probing the evolution
of the ``equatorial dipole'' \citep[e.g.,][]{Wang&Sheeley_1991} in addition to the
``axial dipole'' associated with the $m=0$ modes.  The strength of this
equatorial dipole, and the tipping of the magnetic poles away from the
geographic poles, are indications of the non-axisymmetric contributions to
the magnetism.  

A quantitative assessment of non-axisymmetry
is provided by Figure~\ref{fig:pole wander}$b$, which shows the magnitude of
the $m=0$ component of the dipole and the amplitude of the $m=\pm 1$
dipole components added in quadrature during a roughly 2000-day
interval around this reversal (with second reversal occuring after day
5500).  As the mean toroidal field in the
wreathes reverses polarity (Fig.~\ref{fig:case_D5_reversal}), the
axial dipole diminishes in strength and changes sign, while the
non-axisymmetric pieces fluctuate erratically and do not
seem to sense the global-scale reversal.

The erratic evolution of the non-axisymmetric field components, 
as compared to the relatively smooth evolution of the $m=0$
field, suggests that the dominant processes generating non-axisymmetric
fields are somewhat different from those responsible for axisymmetric
fields.  Both the ``wandering of the poles'' in longitude (Fig.~\ref{fig:pole wander}$a$) and
the fluctuating $m = \pm 1$ field amplitudes (Fig.~\ref{fig:pole wander}$b$) would be expected if
non-axisymmetric field generation were associated mainly with small-scale
magnetic features, that add with random phases to yield a small (but non-zero)
contribution to the dipole.  The longitude of the pole might then be
expected to undergo a random walk on timescales comparable to the
convective eddy turnover time.  This is consistent with what we
observe in case~D5.  

The smooth latitudinal migration of the magnetic pole, on the other
hand, suggests that the global-scale field reversals that occur in
case~D5 rely on more than just the random agglomeration of
uncorrelated small-scale magnetic elements.  Thus the global-scale
axisymmetric $m=0$ fields evolve in a fashion distinct from the
non-axisymmetric components.  We note that when we try to assess the
evolution of the dipole field in the fashion shown here at the
mid-convection zone or near the stellar surface, we find that the
dipole there is much more variable in time.   At these higher radial levels 
the dipole shows several reversals which do not correspond to the 
global-scale reversals of toroidal and poloidal polarity.  Instead in the upper
convection zone, the higher-order modes of the poloidal field more
accurately track the reversals occurring throughout the convection
zone.  We consider these in Section~\ref{sec:higher order moments}.

\subsection{Axisymmetric Dipole, Quadrupole and Octupole}
\label{sec:higher order moments}

The temporal evolution of the axisymmetric dipole, quadrupole, and octupole
moments of the poloidal field can be assessed by measuring the
amplitudes of the $m=0$ components of the radial magnetic field
$B_r$. These amplitudes are defined here as 
\begin{eqnarray}
   \scrD &=& \sqrt{\frac{3}{4 \pi}} 
                        \int{ B_r \cos{\theta}  \sin{\theta} \mathrm{d} \theta \mathrm{d} \phi} \\
   \scrQ &=& \frac{1}{2} \sqrt{\frac{5}{4 \pi}} 
                        \int{B_r (3\cos^2{\theta} - 1) \sin{\theta} \mathrm{d} \theta \mathrm{d} \phi} \\
   \scrO &=& \frac{1}{2} \sqrt{\frac{7}{4 \pi}} 
                        \int{B_r (5\cos^3{\theta} - 3 \cos{\theta})  \sin{\theta} \mathrm{d} \theta \mathrm{d} \phi} 
\end{eqnarray}
where the integral solid angle is  at fixed radius \citep[e.g.,][]{Arfken&Weber_1995}.  
The quantity $\scrD$ is often termed the ``axial dipole'' in solar physics 
\citep[e.g.,][]{Wang&Sheeley_1991}.

\begin{figure}[t]
  \includegraphics[width=\linewidth]{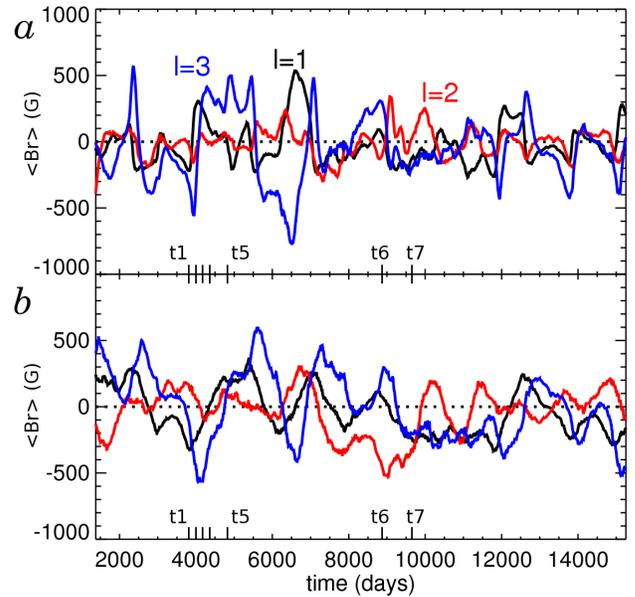}
 \caption{
Temporal evolution of low-degree moments of the poloidal field.
Shown are the axisymmetric ($m=0$) components of the dipole,
quadrupole and octupole moments ($\scrD$, $\scrQ$, and $\scrO$, with
$\ell=1,2,3$ respectively) of the radial magnetic field $B_r$ at two depths: ($a$)
near the surface at $0.97\:R_\odot$ and ($b$) near the base of the
convection zone at $0.73\:R_\odot$.  At depth, the octupole moment
consistently lags the dipole.  During the interval roughly spanning
days 7300--10200 the dynamo enters an unusual symmetric polarity state
and the quadrupole state becomes strong and does not reverse.  After
this period, normal cyclic behavior resumes.
\label{fig:poloidal moments}}
\end{figure}

The amplitudes $\scrD$, $\scrQ$, and $\scrO$ (the $\ell=1,2,3$
components of the $m=0$ field) are shown over an extended
interval of about 13,000 days in Figure~\ref{fig:poloidal moments}.
Here, the measurements are integrated over two spherical surfaces, with
one near the top of the convection zone (Fig.~\ref{fig:poloidal moments}$a$, at
$0.97 R_\odot$) and another near the base of the convection zone
(Fig.~\ref{fig:poloidal moments}$b$, at $0.73 R_\odot$). 
The octupole moment is generally the strongest, particularly when the 
oppositely-directed wreathes are most prominent, but the dipole and
quadrupole moments are also significant.  The typical amplitude of these multipoles 
is comparable to that realized in simulations of solar convection with
large-scale fields \citep{Browning_et_al_2006}, and stronger
than in simulations of small-scale dynamo action by solar convection with
potential field lower boundary conditions \citep{Brun_et_al_2004}.  They are also
substantially stronger than the surface dipole moment observed on the Sun
\citep[e.g.,][]{Wang&Sheeley_1991, Schrijver&DeRosa_2003}.  Peak pointwise values
of the 3D radial field $B_r$ typically exceed 1~kG.

Interpreting the evolution of the axisymmetric poloidal field is made challenging
by the nontrivial depth dependence that the various moments exhibit.  
Near the top of the convection zone, the dipole moment $\scrD$ changes
sign about twenty times over the 13000 days sampled, although
some excursions to one polarity or the other are short-lived
(Fig.~\ref{fig:poloidal moments}$a$).  These 
polarity reversals are separated by fairly irregular intervals of 300 to
1500 days.  The sign of the axisymmetric quadrupole component $\scrQ$ flips
somewhat more frequently, undergoing a few reversals that are not reflected
in the dipole moment.  Generally the dipole and quadrupole track each
other well in time, with no consistent lag between the two moments.
The octupole reverses less frequently than either the dipole or
quadrupole, and when it does reverse it generally lags them slightly in time.

In the lower convection zone (Fig.~\ref{fig:poloidal moments}$b$), the
moments evolve relatively slowly, with fewer reversals separated by
significantly longer intervals of 
about 1500 days.  The octupole moment consistently lags a few hundred days behind the dipole.
The longer evolution times in the lower convection zone
could plausibly reflect the longer Alfv\'en times deeper down
or the longer magnetic diffusion time -- for our choice of
SGS magnetic diffusivity $\eta$, scaling with $\bar{\rho}^{-1/2}$,
both timescales vary in roughly the same fashion with depth.
Convective time scales also decrease toward the bottom of the convection
zone and the ratio of magnetic to kinetic energy increases.  Thus,
the more stable mean fields may reflect less buffeting by convective
motions.

The temporal variations in these low-degree moments are linked to the
reversals of toroidal field, but the relationship between the
different field components is complex.  Near the surface, the octupole
tracks the deep-seated toroidal fields fairly well, but the dipole and
quadrupole moments exhibit flips which do not correspond to
global-scale reversals.  In the lower convection zone, it appears that
the dipolar reversal precedes the octupolar reversals.  Typically the
reversal of the dipole occurs shortly after the octupole is near its
peak amplitude (e.g., at about time~t4).  The octupole then generally
reverses when the dipolar moment is near a maximum in amplitude (e.g.,
about time~t5).  Most of the reversals in dipole and octupole polarity
deep in the lower convection zone correspond to a flip in the sign of
the predominant toroidal field in each hemisphere, but typically occur
only after the wreaths have reversed their polarities in the near
equatorial region (this occurs by time~t3 in our example reversal).

When the dipole and octupole are strong, we generally see wreaths that
are antisymmetric about the equator.  When the quadrupole becomes
stronger (e.g., during the interval from roughly 7000--10000 days) the
dynamo typically exhibits toroidal fields that are symmetric about the
equator.  This indicates that the global-scale toroidal field is
likely being produced by the stretching of global-scale poloidal field
by the shear of differential rotation.

\begin{figure*}
  \begin{center}
  \includegraphics{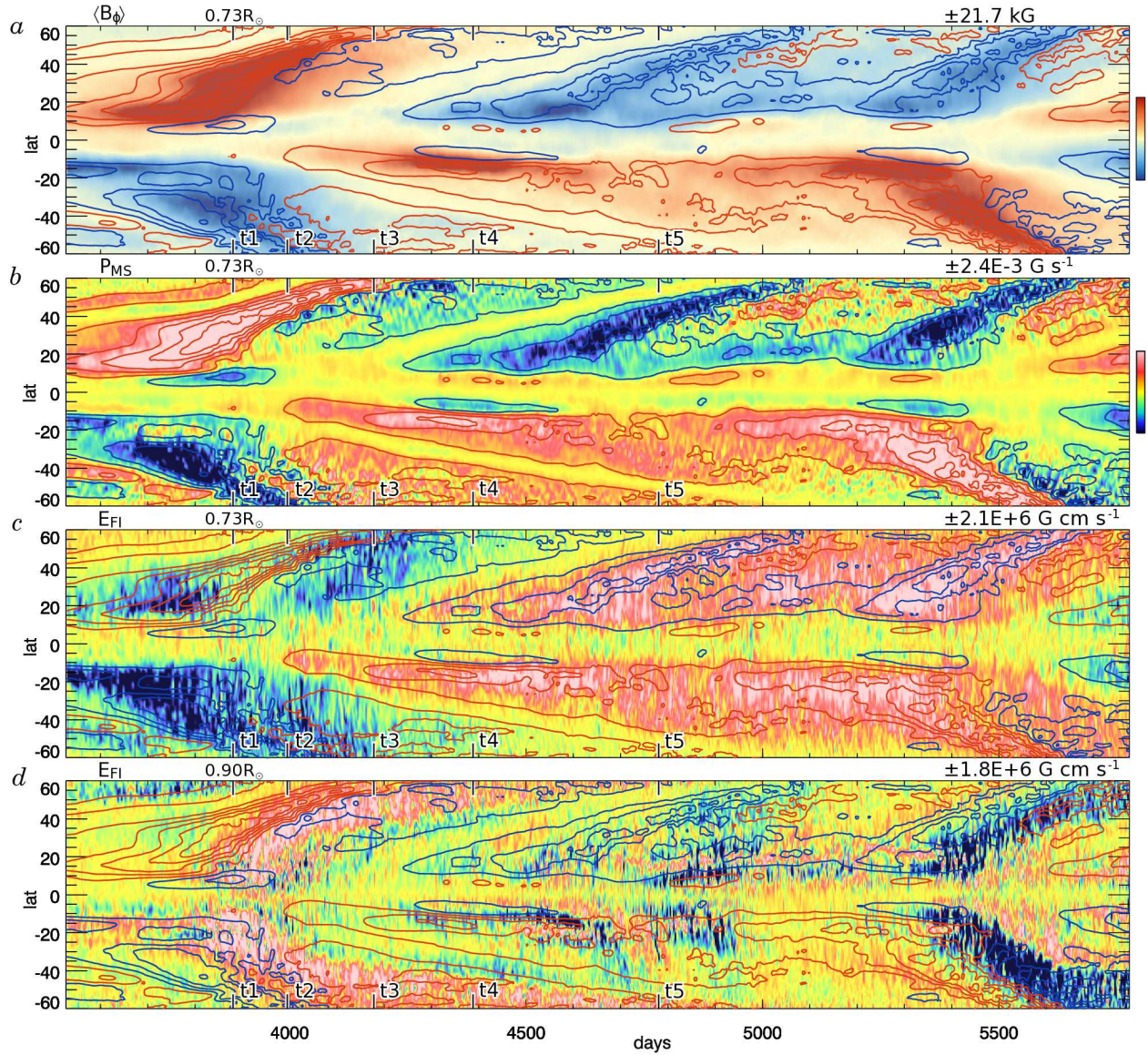}
  \end{center}
  \vspace{0.1cm}
  \caption{Temporal and latitudinal variation of magnetic fields and the effects that
    generate them.  ($a$) Longitudinally-averaged toroidal field
    $\langle B_{\phi} \rangle$ near the base of the convection zone
    ($0.73R_\odot$), as image and contour plot in latitude and time.
    ($b$) Production of toroidal fields by mean shear
    $P_\mathrm{MS}$.
    ($c$) Longitude-averaged turbulent electromotive force
    $E_\mathrm{FI}$ which generates poloidal vector potential $\langle
    A_\phi \rangle$.  
    ($d$) Same emf $E_\mathrm{FI}$ shown in the upper convection zone
    ($0.90R_\odot$).  
    Contours of $P_\mathrm{MS}$ at $0.73 R_\odot$
    are overplotted in all panels for reference (red positive, blue
    negative).  These contours have been smoothed with a 35-day moving
    time average.
    \label{fig:D5 production}}
\end{figure*}

\section{A Nonlinear Dynamo Wave}
\label{sec:D5 production}

The latitudinal propagation and orderly reversal of large-scale magnetic
fields realized in case D5 constitute a striking finding of these
simulations.  Here we examine how field propagation and reversal might
arise, by examining the spatial and temporal dependence of some of the
mechanisms that act to strengthen or weaken fields.

Spatial propagation and field reversals may be expected whenever the
processes that amplify or reduce a magnetic field -- i.e., source
terms in the induction equation -- are out of phase spatially or
temporally with the field itself.  In mean-field solar dynamo models,
such spatial propagation can arise through the combined operation of
the $\alpha$-effect and $\Omega$-effect in the form of a 
``dynamo wave'' \citep[e.g.,][]{Stix_1976, Yoshimura_1976}.  In mean-field
solar dynamo models that employ a spatial separation between poloidal
and toroidal source regions, flux transport by the meridional
circulation, turbulent pumping, or turbulent diffusion can produce
well-defined, non-local, phase relationships between the resulting
mean fields.  For example, in many Babcock-Leighton and interface
dynamo models the poloidal field at the surface is closely linked
to the the toroidal field near the base of the convection zone at a
previous time
\citep[e.g.,][]{Charbonneau&MacGregor_1997,Jouve_et_al_2010}.

In \mbox{Paper I}, we noted that the generation of mean poloidal and 
toroidal field by turbulent fluctuations is not well represented 
by a simple scalar $\alpha$-effect.  This conclusion also applies
to the simulation reported here.  Yet, assessing the
spatial and temporal phasing between mean poloidal and toroidal fields
and their source terms is nevertheless essential in order to understand
the physical mechanisms underlying the spatial propagation and
cyclic reversals exhibited by case D5.  

As in case D3 of \mbox{Paper I}, the principal source of mean toroidal
field in case D5 is the $\Omega$-effect, the conversion and
amplification of mean poloidal field by the mean shear.  In the
language of \mbox{Paper I}, this production term is
\begin{multline}\label{Tsource}
  P_\mathrm{MS} = \left( \langle \vec{B} \rangle \cdot \vec{\nabla}\right)\langle \vec{v}\rangle|_\phi =\\ \advbm  \langle v_{\phi} \rangle  + \frac{ \langle B_{\phi} \rangle  \langle v_r \rangle  + \cot\theta  \langle B_{\phi} \rangle  \langle v_{\theta} \rangle }{r} ~,
\end{multline}
and the principle source of mean poloidal field is the longitudinal
component of the turbulent emf: 
\begin{equation}\label{Psource}
  E_\mathrm{FI} = \langle \vec{v}^\prime \times \vec{B}^\prime \rangle |_\phi~.
\end{equation}
The curl of $E_\mathrm{FI} \vec{\hat{\phi}}$ contributes to the time
derivative of $\langle \vec{B}_\mathrm{pol} \rangle$ (cf.\ eq.\
\ref{Bpol}).  As in previous sections, angular brackets represent
averages over longitude and primes denote fluctuations about the mean,
e.g.\ $\vec{v}^\prime = \vec{v} - \langle \vec{v}\rangle$.

The dynamical balances described in detail in \mbox{Paper I} for case
D3 largely apply also to case D5.  The source terms (\ref{Tsource})
and (\ref{Psource}) are opposed by ohmic diffusion and by the
meridional components of the turbulent emf which act to fragment and
disperse the mean toroidal field.  The dispersal of the mean toroidal
field by convective motions encompasses the concepts of turbulent
diffusion and magnetic pumping but it is generally more complex, with
intricate 3D structure and subtle nonlinear feedbacks.
Advection by the meridional circulation can also contribute to the
time evolution of the mean toroidal and poloidal fields but we find
that its role in these simulations is relatively minor.  This is in
stark contrast to Flux-Transport solar dynamo models where the
advection of toroidal flux by the meridional circulation regulates
cyclic activity
\citep[e.g.,][]{Wang&Sheeley_1991,Dikpati&Charbonneau_1999,Kuker_et_al_2001,Nandy&Choudhuri_2001,Dikpati&Gilman_2006,Rempel_2006,Jouve&Brun_2007,Yeates_et_al_2008}. In case D5, it is repetitive, systematic imbalances 
between these multiple production and dissipation terms which, unlike 
in case D3, gives rise to cyclic behavior.

Figure~\ref{fig:D5 production}$a$ exhibits the the mean toroidal field
$\langle B_{\phi} \rangle$  together with ($b$) 
the principle toroidal and ($c$) poloidal source terms in the lower
convection zone 
(at $0.73R_\odot$) over an interval of about 2300 days, spanning one full
reversal of the global-scale field.  The poloidal source term
$E_\mathrm{FI}$ is also shown in the upper convection zone
(Fig.~\ref{fig:D5 production}$d$, at $0.90R_\odot$) and contours of
$P_\mathrm{MS}$ are shown over each quantity, to aid in assessing
phase relationships.

It is apparent from Figures~\ref{fig:D5 production}$a,b$ that
$\langle B_\phi \rangle$ and $P_\mathrm{MS}$ trace one another closely
and largely possess the same sign, highlighting the role of rotational
shear in generating and maintaining the wreaths.  Closer scrutiny
reveals more intricate phase relationships.  The $\Omega$-effect
precedes the appearance of the wreaths, confirming its role as
the principle source term.  Furthermore, it is skewed poleward
relative to the main toroidal flux concentration.  This,
coupled with a subsequent reversal in sign of $P_\mathrm{MS}$ on 
the equatorward edge of the wreaths, induces poleward 
propagation.  

Thus the $\Omega$-effect amplifies the poleward edge of the wreath
while simultaneously suppressing the equatorward edge.  The mean
poloidal field is generated in the vicinity of the wreaths, moving
poleward in conjunction with the toroidal flux and the toroidal source
term $P_\mathrm{MS}$ Figure~\ref{fig:D5 production}$c$.  Thus, the
poleward propagation may be regarded as a nonlinear manifestation of a
dynamo wave.  It is nonlinear in the sense that the poloidal source
term $E_\mathrm{FI}$ is not linearly proportional to the mean field
and it is a dynamo wave in the sense that spatial propagation is
induced by means of the relative phasing of poloidal and toroidal
source terms.  Over time, this reinforcement at higher latitudes 
and cancellation at low latitudes contributes to the poleward
propagation and ultimately to the reversal.  Dynamo waves like this
appear to have been captured in early solar dynamo simulations by
\cite{Gilman_1983} and \cite{Glatzmaier_1985}.

Generally the turbulent emf $E_\mathrm{FI}$ (Fig.~\ref{fig:D5
production}$c$) tracks the mean toroidal field in time and in space,
though its sign is symmetric about the equator while $\langle B_\phi
\rangle$ is generally antisymmetric.  Closer scrutiny reveals a slight
offset toward the poleward side of the wreaths where the
$\Omega$-effect also operates.  Thus, in contrast to traditional
$\alpha$-$\Omega$ dynamos, a systematic phase shift between 
$\langle B_{\phi} \rangle$ and $E_\mathrm{FI}$ appears to
contribute to the latitudinal propagation.

Figure~\ref{fig:D5 production}$d$ indicates that the poloidal field in
the upper convection zone is established largely in response to field
generation in the lower convection zone.  In particular,
$E_\mathrm{FI}$ does not begin changing in the upper convection zone
until the fields in the deep convection zone are already in the
process of reversing (times t1--t2).  This term continually alters the
mean poloidal field in the upper convection zone relatively unimpeded,
in contrast to the lower convection zone where it is almost entirely
balanced by ohmic diffusion.

In order to fully characterize the role of the source terms in
promoting polarity reversals and poleward propagation, we must
understand them within the context of the mean poloidal field
evolution shown in Figure~\ref{fig:A_phi}.  The dominant contribution
to $P_\mathrm{MS}$ is from the $\Omega$-effect, proportional to
$\langle \vec{B}_\mathrm{pol} \rangle \vec{\cdot \nabla} \Omega$.
Since the contours of $\Omega$ are primarily cylindrical, 
$\vec{\nabla} \Omega$ is predominantly directed away from
the rotation axis (Fig.~\ref{fig:case_D5_patterns}).  It is then this
component of the mean poloidal field we must consider when following
the evolution of $P_\mathrm{MS}$, namely 
$\langle B_\mathrm{s} \rangle = \langle \vec{B}_\mathrm{pol} \rangle \vec{\cdot
\nabla} s$ where $s = r \sin\theta$ is the moment arm.

The time span between t1 through t5 provides a demonstration of the
processes involved.  At time t1, the predominantly octupolar
configuration of the mean poloidal field (Fig.~\ref{fig:A_phi}$a$)
coupled with the perfectly conducting lower boundary condition yield a
positive $\langle B_\mathrm{s} \rangle$ (away from the rotation axis)
through much of the northern hemisphere.  This reverses near the
equator where $\langle B_\mathrm{s} \rangle$ is directed toward the
rotation axis.  The associated transition between positive and
negative $\langle A_\phi \rangle$ occurs across the wreaths.
Rotational shear operating on this poloidal field structure through
the $\Omega$-effect accounts for the positive sign of $P_\mathrm{MS}$
through much of the northern hemisphere at time t1 (Figure~\ref{fig:D5
production}$b$) as well as the poleward skewness and the low-latitude
sign reversal.  As time proceeds, the neutral surface $\langle A_\phi
\rangle = 0$ drifts poleward along with the wreaths and gradually
evolves from a radial orientation to a more horizontal orientation by
t3 (Fig.~\ref{fig:A_phi}$c$).   This is associated with the opening up
of the poloidal field in the upper convection zone noted in 
Section~\ref{sec:poloidal reversals}, as low-latitude loops spread 
poleward.  The net result is a reversal in $\langle
B_\mathrm{s} \rangle$ (from positive to negative at mid-latitudes
in the northern hemisphere), with a corresponding reversal in
$P_\mathrm{MS}$ (Figure~\ref{fig:D5 production}$b$).

Rotational shear operating on the cylindrically inward poloidal field
at the equator at time t3 soon produces toroidal field of the opposite
sign (negative in the northern hemisphere).  By t4, the low-latitude
wreaths dominate the toroidal field structure and the reversal is
complete (Fig.~\ref{fig:case_D5_reversal}$e$).  
Near the equator, $E_\mathrm{FI}$ begins to reverse sign in the lower
convection zone around time~t2 but remains weak while the wreaths move poleward.
As the new low-latitude toroidal wreaths become established between t3 and t4,
the reversed $E_\mathrm{FI}$ grows in amplitude near the equator,
generating mean poloidal field of the opposite sense 
relative to higher latitudes and previous times
(Fig.~\ref{fig:A_phi}$d$). This transition is likely a culmination of
the evolving 3D (non-axisymmetric) magnetic linkages across the
equatorial plane highlighted in Section~\ref{sec:xeq}.  Subsequent evolution
enhances the octupolar structure of the mean poloidal field
(Fig.~\ref{fig:A_phi}$e$), setting the stage for the next reversal.

Thus the helical nature of the wreaths promotes their poleward
propagation and subsequent polarity reversal.  This is not to say
that the mean fields are simply twisted tori.  Rather, the maxima and
minima of $\langle A_\phi \rangle$ are displaced relative to $\langle
B_\phi \rangle$ such that the local magnetic helicity density of the
mean field $\langle A_\phi \rangle \langle B_\phi \rangle$ changes
sign across each wreath.  Yet the resulting magnetic topology
exhibits regions of oppositely-directed $\langle B_\mathrm{s} \rangle$
near the wreath's poleward and equatorward edges.  This induces
poleward propagation by means of a sign reversal in $P_\mathrm{MS}$
as described above.

A potential alternative interpretation of wreath evolution in case D5
is the poleward slip instability whereby axisymmetric rings of
toroidal flux drift poleward as a consequence of magnetic tension
\citep[e.g.,][]{Spruit&vanBallegooijen_1982,
Moreno-Insertis_et_al_1992}.  However, if this were occurring then the
poleward propagation of $\langle B_\phi \rangle$ would be achieved by
means of a meridional circulation induced by the Lorentz force.
Instead, we find that the amplitude and structure of this advective
contribution to the mean magnetic induction are not sufficient to
account for the observed evolution of $\langle B_\phi \rangle$,
particularly with regard to the poleward propagation.

\section{Torsional Oscillations}
\label{sec:torsional oscillations}

The strong magnetic fields achieved in case~D5 couple strongly with
the global-scale flow of differential rotation.
As the fields themselves vary in strength, the differential rotation
responds in turn, becoming stronger as the fields weaken and then
diminishing as the fields are amplified.  These cycles of faster and
slower differential rotation are visible in the traces of DRKE shown
previously in Figure~\ref{fig:case_D5_energies}$a$.  
We revisit here the interval explored in close detail in 
Sections~\ref{sec:toroidal reversals} and \ref{sec:D5 production},
spanning days 3500 to 5700 of the simulation and one full polarity
reversal.   

\begin{figure}
  \begin{center}
    \includegraphics[width=\linewidth]{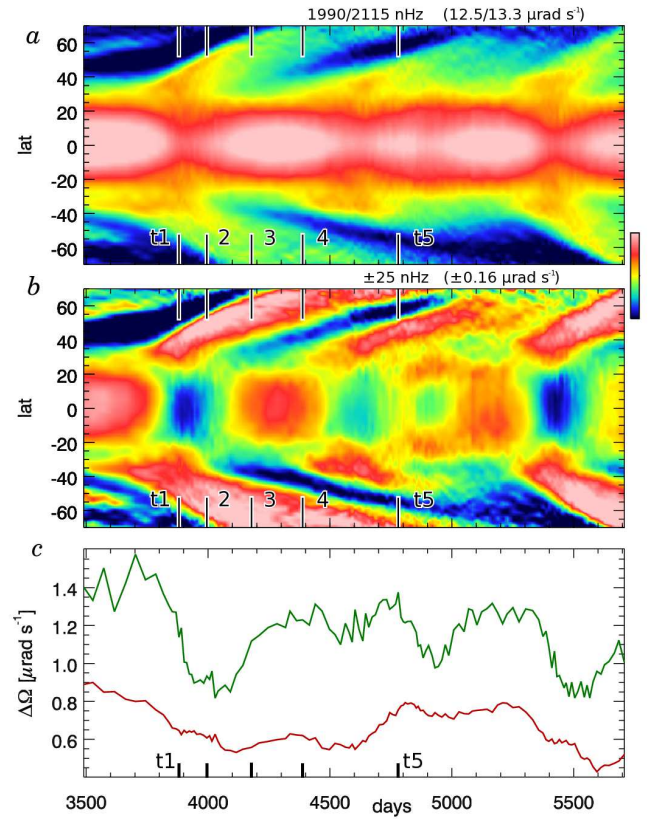}
  \end{center}
  \caption[Time-varying differential rotation in case~D5] 
	  {Time-varying differential rotation in case~D5. 
    $(a)$~Time-latitude map of angular velocity $\Omega$ at
    mid-convection zone ($0.85R_\odot$).  There are substantial
    temporal variations at both the equator and high latitudes.  
    $(b)$~These are accentuated by subtracting the time-averaged
    profile of $\Omega(r,\theta)$ at each latitude.  Visible are
    poleward propagating speedup structures at high latitudes and more
    uniform modulations near the equator.
    $(c)$~Corresponding variations in $\Delta \Omega_\mathrm{lat}$
    near the surface (upper~curve, green) and at mid-convection zone
    (lower, red).
  \label{fig:case_D5_omega}}
\end{figure}

\subsection{Bands of Shear}

The angular velocity $\Omega$ at mid-convection zone is shown for this
period as a time-latitude map in Figure~\ref{fig:case_D5_omega}$a$.
Here again the timing marks indicate our fiducial times~t1--t5.
In the equatorial regions, the differential rotation remains fast and
prograde, but with some modulation in time.  Prominent structures of
speedup are visible propagating toward the poles at the high
latitudes.   These structures are much more evident when we subtract
the time-averaged profile of $\Omega$ for this period at each
latitude (Figure~\ref{fig:case_D5_omega}$b$).  These bands appear as strong,
tilted fast (red) bands extending poleward from roughly $\pm30^\circ$
latitude.  In the northern hemisphere, three such speedup bands
are launched over this interval.  In contrast, in the south only two
such bands are evident; a third is perhaps launched around day 4500,
but it does not survive or propagate.  

Comparing these features with the propagation of magnetic fields shown in
Figures~\ref{fig:case_D5_reversal}$a$ and 
\ref{fig:D5 production}$a$ over the same interval, we find
that velocity speedup features are well correlated with the poleward
migration of mean longitudinal magnetic field.
The velocity features bear some resemblance to the poleward
branch of torsional oscillations observed in the solar convection
zone over the course of a solar magnetic activity cycle, though on a
much shorter time scale here as befits the correspondingly shorter time
between magnetic polarity reversals in these dynamo simulations.  

The speedup bands propagate toward the poles relatively
slowly.  In a period of roughly 500 days they travel about $40^\circ$
in latitude, and this angular propagation rate of about 
$0.08 ^\circ\:\mathrm{day}^{-1}$ or about 
$1.6 \times 10^{-8} \mathrm{rad}\: \mathrm{s}^{-1}$
is roughly constant at all depths in the convection zone. 
At mid-convection zone ($0.85 R_\odot$) this corresponds to a distance
of roughly 415~Mm and a propagation velocity of about
$0.8~\mathrm{Mm}\thinspace\thinspace\mathrm{day}^{-1}$ or about $9~\ms$. 
Near the base of the convection zone ($0.73R_\odot$), this corresponds
to a distance of about 355~Mm and a propagation velocity of 
$0.7~\mathrm{Mm}\thinspace\thinspace\mathrm{day}^{-1}$ or about $8~\ms$.

This is considerably slower than the fluctuating latitudinal flows
associated with the convection which at mid-convection zone have peak
speeds of $\pm 200~\ms$ during this time period.  The meridional
circulations are more difficult to interpret.  At mid-convection zone
they can have instantaneous amplitudes of about $\pm 10~\ms$, but the
fluctuations are large and the time-averaged circulations are much
slower ($\sim 1$--$2~\ms$) though they are poleward in sense at
mid-latitudes.  Near the base of the convection zone the time-averaged
circulations are weakly equatorward with amplitudes of about 
$1~\ms$ and thus act to resist the poleward propagation of the speedup
bands and the magnetic wreaths.

If we do moving 35-day time-averages of the meridional flows in the
lower convection zone, we see some evidence for variations that track
the speedup bands.  In particular, the flow in the core of the
wreathes tends to be poleward with an amplitude of up to $1$--$3~\ms$
(up to $5~\ms$ without smoothing).  This poleward flow is likely
induced by magnetic tension in the toroidal wreathes, analogous to the
mechanism underlying the poleward slip instability
\citep[e.g.,][]{Moreno-Insertis_et_al_1992,Jouve&Brun_2009}. In the
absence of rotation and convection, the propagation speed
associated with the polar slip instability is approximately given by
the Alfv\'en speed associated with the mean toroidal fields,
\begin{equation} 
  v_{A,\phi} = \frac{\langle B_\phi \rangle}{\sqrt{4 \pi \bar{\rho}}}.  
\end{equation} 
In case D5, the propagating toroidal
features have an amplitude of $\langle B_\phi \rangle \sim$
$3$--$6$~kG in the mid convection zone, implying an Alfv\'en speed of
about $30$--$60~\ms$.  Toroidal field strengths are larger near the
base of the convection zone, $15$--$25$~kG, yielding larger Alfv\'en
speeds, $\sim 100$--$160~\ms$, despite the higher density
($0.19~\mathrm{g}\thinspace\thinspace\mathrm{cm}^{-3}$, as opposed to
$0.065~\mathrm{g}\thinspace\thinspace\mathrm{cm}^{-3}$ in the mid
convection zone).  The much lower poleward flows in the wreathes may
reflect the inhibiting influence of rotation
\citep{Moreno-Insertis_et_al_1992} and convective pumping.

In the lower convection zone, the slip-induced poleward flow must also
operate against the background equatorward flow maintained by the
convection (Fig.~\ref{fig:case_D5_patterns}$e$).  This gives rise to a
horizontal convergence of the meridional flow near the poleward edge
of each wreath and an associated upward flow with an amplitude of
about $1~\ms$, consistent with mass conservation.  This is analogous
to the recirculating flows around shear-generated magnetic flux
structures that rise due to magnetic buoyancy
\cite[e.g.,][]{Cline_et_al_2003a,Cline_et_al_2003b}.

In the idealized polar slip instability, the poleward propagation of
toroidal bands is associated with the formation of prograde zonal
flows as the fluid within the bands tends to conserve its angular
momentum.  By contrast, the wreathes in case D5 are not isolated flux
structures, they have a leaky topology that allows plasma to escape.
This will also influence the slip-induced propagation speed but it is
likely to accelerate it rather than decelerate it because prograde
zonal flows provide gyroscopic stabilization.  The wreathes in case D5
do induce prograde zonal flow variations as they migrate poleward
(Fig.~\ref{fig:case_D5_omega}), as expected from the polar slip
instability.  However, as we shall see in Section~\ref{sec:amom transport}, the
zonal flow variations in case D5 are largely produced by the Lorentz
force, as opposed to the Coriolis-induced flows associated with
angular momentum conservation in closed flux structures.

The important role of the mean Lorentz force in accelerating the
torsional oscillations (Section~\ref{sec:amom transport}) and the phasing of the
$\Omega$-effect relative to the wreathes (Section~\ref{sec:D5 production})
suggests that the poleward propagation of the torsional oscillations
may be mainly attributed to a nonlinear dynamo wave.  The phase speed
of such a wave is governed primarily by the poloidal and toroidal
source terms in the mean induction equation but it may also be linked
to the Alfv\'en speed associated with the mean latitudinal field.
Values of $\langle B_\theta \rangle$ in the cores of the poleward
propagating bands are roughly $2$--$3$~kG in the mid convection zone,
implying Alfv\'en speeds of about $20$--$30~\ms$.  Near the base of
the convection zone these values are larger, with mean latitudinal
field strengths of $6$--$9$~kG and Alfv\'en speeds of $40$--$60~\ms$.

With the expanded sensitivity of Figure~\ref{fig:case_D5_omega}$b$, 
we can see that the equatorial modulation appears as fast and slow
pulses which span the latitude range of $\pm20^\circ$.  These
variations are fairly uniform across this equatorial region.  
The velocity variations at the equator do not correspond with the
equatorial propagating branch of torsional oscillations seen in the
Sun \citep{Thompson_et_al_2003}.  In the Sun, the equatorial branch
may arise from enhanced cooling in the magnetically active regions
\citep[e.g.,][]{Spruit_2003, Rempel_2006, Rempel_2007}.

\begin{figure*}[!t]
  \begin{center}
    \includegraphics[width=\linewidth]{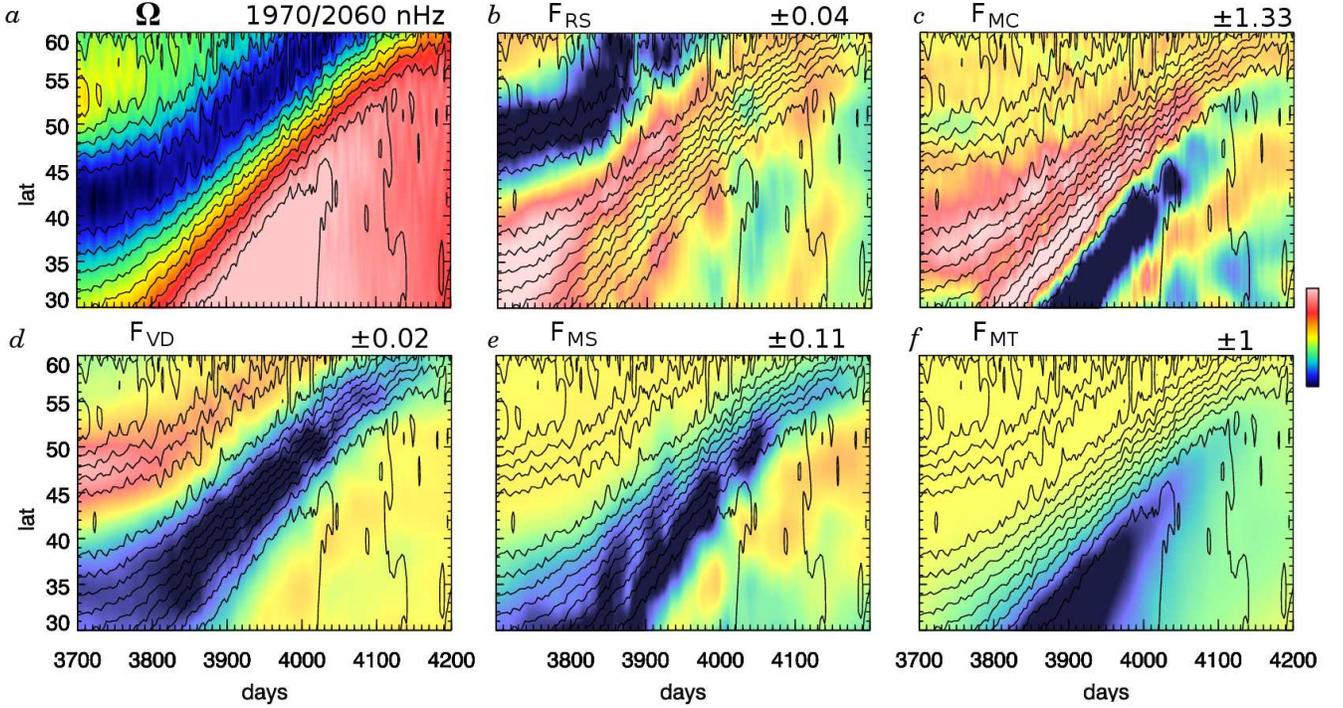}
  \end{center}
\caption{Angular momentum transport in case~D5 in the northern
  hemisphere of the lower convection zone ($0.73R_\odot$).  
  $(a)$ Angular velocity $\Omega$ as image and contour plot in
  latitude and time.  Shown at the same depth, with contours of
  $\Omega$ overlain to guide comparison, are latitudinal angular momentum fluxes
  arising from 
  $(b)$ Reynolds stresses ($F_\mathrm{RS}$),
  $(c)$ meridional circulations ($F_\mathrm{MC}$),
  $(d)$ viscous diffusion ($F_\mathrm{VD}$),
  $(e)$ Maxwell stresses from fluctuating fields  ($F_\mathrm{MS}$), and
  $(f)$ magnetic torques from axisymmetric fields ($F_\mathrm{MT}$).
  Indicated amplitudes of fluxes are all scaled relative to
  $F_\mathrm{MT}$, with negative (positive) fluxes in the northern
  (southern) hemisphere representing poleward transport.  All fluxes
  ($b-f$) have been smoothed with a 35-day moving time-average, but
  the contours of $\Omega$, with no temporal smoothing, give a sense of the
  variations.
  \label{fig:D5 amom}}
\end{figure*}

The temporal variations of the angular velocity contrast in latitude
$\Delta \Omega_\mathrm{lat}$ is shown for this period in
Figure~\ref{fig:case_D5_omega}$c$.  At mid-convection zone (sampled by
red line) the variations in $\Delta \Omega_\mathrm{lat}$ can be
substantial,  with large contrasts when the fields are
strong in the magnetic cycle (prior to t1) and smaller contrasts when the
fields are in the process of reversing (t2, t3). 
Near the surface (green line) $\Delta \Omega_\mathrm{lat}$ can show
even larger variations.  These near-surface
values of $\Delta \Omega_\mathrm{lat}$ are reported in
Table~\ref{table:delta_omega}, averaged over this entire period
(avg) and at points in time when the contrast is large (max,
at day~3702) and small (min, at day~4060).  From periods of highest
contrast to lowest, $\Delta \Omega$ changes by about 
0.5~$\mu  \mathrm{rad}\: \mathrm{s}^{-1}$, which is a change of roughly
45\% relative to the long running average.  This corresponds to a
total change of about 4\% relative to the frame rotation rate of 
13~$\mu  \mathrm{rad}\: \mathrm{s}^{-1}$ ($5\: \Omega_\odot$).

\subsection{Angular Momentum Transport}
\label{sec:amom transport}

We now turn to discussing the transport of angular momentum associated
with the speedup bands observed in case~D5. The bands of speedup are
clearly associated with the magnetic wreaths which are similarly
propagating towards the poles, and here we examine the physical
processes that lead to the angular momentum transport that locally
speeds up the flow of differential rotation.

Our choice of stress-free and potential-field/perfectly conducting
boundary conditions at the top and the bottom of the shell
respectively has the advantage that no net external torque is applied
and angular momentum is conserved.  Convection and magnetism can
however redistribute angular momentum throughout the shell.  
In the general MHD case there are five processes that serve to
transport angular momentum \citep{Brun_et_al_2004}.  
These are the Reynolds stresses from fluctuating flows (which we
denote $F_\mathrm{RS}$), the meridional circulations 
($F_\mathrm{MC}$), the viscous torque ($F_\mathrm{VD}$), the Maxwell stresses from
fluctuating fields ($F_\mathrm{MS}$), and the magnetic torque from the
global-scale fields ($F_\mathrm{MT}$). 
The latitudinal component of the angular momentum flux $\cal{F}$ is thus:
\begin{equation}
{\cal{F}}_{\theta}= F_\mathrm{RS} + F_\mathrm{MC} + F_\mathrm{VD} +F_\mathrm{MS} + F_\mathrm{MT},
\end{equation}
with
\begin{eqnarray}
  F_\mathrm{RS} &=& \bar{\rho} r \sin\theta \langle v_{\theta}^{'} v_{\phi}^{'} \rangle,\\
  F_\mathrm{MC} &=& \bar{\rho} r \sin\theta \langle v_{\theta}\rangle(\langle {v}_{\phi} \rangle+\Omega_0 r\sin\theta),\label{eq:F_MC}\\
  F_\mathrm{VD} &=&-\nu \bar{\rho} \sin^2\theta \frac{\partial}{\partial \theta}\left(\frac{\langle v_{\phi} \rangle}{\sin\theta}\right),\\
  F_\mathrm{MS} &=&-\frac{r \sin\theta}{4\pi} \langle B_{\theta}^{'}B_{\phi}^{'} \rangle, \\
  F_\mathrm{MT} &=&-\frac{r \sin\theta}{4\pi} \langle {B}_{\theta}\rangle \langle {B}_{\phi} \rangle.
\end{eqnarray}
These five latitudinal angular momentum fluxes are displayed for
case~D5 in Figure~\ref{fig:D5 amom} along with the angular velocity
itself (Figure~\ref{fig:D5 amom}$a$).  These measurements are
taken in the lower convection zone where the magnetic wreaths are
strong (at $0.73R_\odot$) and as in \cite{Ballot_et_al_2007} we do not
average these contributions in time but rather follow their time
evolution with time-latitude plots.  Here we focus on an interval
spanning 3700--4200 days when the speedup bands are launched during
the reversal.  We show only the northern hemisphere
here, but dynamics in the southern hemisphere are similar.
Contours of angular velocity $\Omega$ are superimposed
on each panel to guide the eye.  

The transport of angular momentum is modulated and has
significant polar branches that are strongly associated with the
speedup bands.  As the magnetic wreaths propagate
poleward, they accelerate the local angular velocity.  This is clearly
seen  in Figure~\ref{fig:D5 amom}$a$, where the latitudinal shear is
reduced with the retrograde (blue) regions disappearing as the
prograde (red) bands migrate polewards. The acceleration of the high
latitude region is largely due to the large-scale magnetic torque
associated with the  axisymmetric fields in the magnetic wreaths
($F_\mathrm{MT}$, Figure~\ref{fig:D5 amom}$f$).  This magnetic torque traces the
propagation of $\langle B_\phi \rangle$ quite well.  The 
negative (positive) sense of this term in the core of the wreaths in
the northern (southern) hemisphere indicates that $F_\mathrm{MT}$
transports angular momentum toward the poles, thus accelerating flow
in those retrograde regions.

The Maxwell stresses from the fluctuating fields ($F_\mathrm{MS}$,
Figure~\ref{fig:D5 amom}$e$) also help move angular momentum polewards
but contribute more weakly than the axisymmetric fields.  In the
equatorial region $F_\mathrm{MS}$ continues to operate even during the
reversal when the axisymmetric fields, and hence $F_\mathrm{MT}$, are
very small. The Reynolds stresses ($F_\mathrm{RS}$, Figure~\ref{fig:D5
  amom}$b$) generally act to transport angular momentum towards the
equator but have amplitudes lower than either of the Maxwell
stresses.  In this simulation angular momentum transport by viscous
diffusion ($F_\mathrm{VD}$, Figure~\ref{fig:D5 amom}$d$) plays very
little role.

The Coriolis forces associated with the meridional circulations works
in concert with the Maxwell torques to spin up the flow in the wreaths
($F_\mathrm{MC}$, Figure~\ref{fig:D5 amom}$c$).  
This spinup largely arises from the poloidal flows associated with the
wreaths themselves, rather than from the time-averaged meridional
circulations, which are equatorward at this depth.
Before the speedup bands begin propagating towards the poles,
$F_\mathrm{MC}$ is large and opposite in sense to $F_\mathrm{MT}$,
acting to transport angular momentum towards the equator.  In the
cores of the wreaths however, $F_\mathrm{MC}$ is converging,
which acts to spinup the fluid within the wreaths.  After the
speedup bands have launched (e.g., at about day 3850 in the northern
hemisphere), $F_\mathrm{MC}$~changes sense and acts to 
help maintain the prograde rotation in the speedup bands.  
As the wreaths propagate towards the poles, 
$F_\mathrm{MC}$ is similar in amplitude to $F_\mathrm{MT}$ and
of the same sense.  This large and changing $F_\mathrm{MC}$ arises
almost entirely from the second term in equation~(\ref{eq:F_MC}); the
first term is much smaller in amplitude (roughly 1\% of $F_\mathrm{MT}$).
When no time-averaging is applied, $F_\mathrm{MC}$ exhibits large
fluctuations on both short and long timescales, with instantaneous
amplitudes five or more times larger than $F_\mathrm{MT}$. 

\subsection{Sampling Many Magnetic Cycles in Case~D5}
\label{sec:long_time_D5}

\begin{figure}
 \begin{center}
  \includegraphics[width=\linewidth]{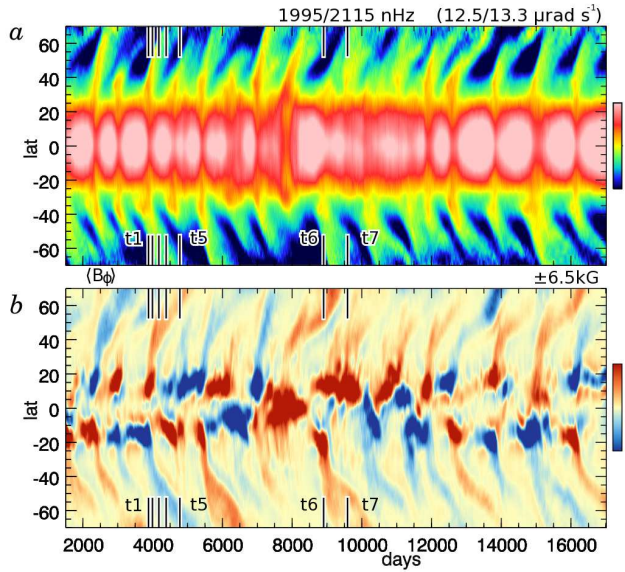}
\end{center}
 \caption[Extended history of varying differential rotation in case~D5]
	  {Extended history of varying differential rotation in case~D5. 
  $(a)$~Variations of $\Omega(r,\theta)$ at mid-convection zone.
  $(b)$~Mean longitudinal field $\langle B_\phi \rangle$, at same depth.
  Poleward propagating speedup
  structures are visible during each magnetic oscillation.
 The time samples used in Figures~\ref{fig:case_D5_reversal}
  and~\ref{fig:case_D5_single_polarity}  are indicated.
\label{fig:case_D5_long_omega}}
\end{figure}

The linked variations of angular velocity and magnetic fields are shown
over considerably longer intervals of time in Figures~\ref{fig:case_D5_long_omega}$a,b$.  
Here too we see the equatorial modulation over many magnetic cycles and the
poleward propagating speedup bands.  Asymmetries between the northern
and southern hemisphere are evident at many times in different
cycles.  The latitudinal angular velocity contrasts exhibit large
variations during each cycle, with $\Delta \Omega_\mathrm{lat}$ near the
surface ranging from 0.6--1.55 $\mu \mathrm{rad}\: \mathrm{s}^{-1}$
during the interval shown here. 
Successive magnetic cycles can have distinctly different angular
velocity contrasts, and there are additional long-term modulations
that span many magnetic cycles.  These effects may be very important
for stellar observations of differential rotation.

These angular velocity variations are consistently associated
with variations in the magnetism.  The mean toroidal field $\langle B_\phi
\rangle$ is shown over the same interval in
Figure~\ref{fig:case_D5_long_omega}$b$.
The poleward propagating magnetic features shown previously in
Figures~\ref{fig:case_D5_reversal}$a$ are evident throughout this
longer time sampling, now appearing as nearly vertical streaks in 
$\langle B_\phi \rangle$, owing to the compressed time axis.
From day 1500 to 7300, four cycles occur in which wreaths of opposite polarity are
achieved in each hemisphere.  After this period, the dynamo explores
unusual single-polarity states.  Here either both wreaths have the
same polarity (t6) or a single dominant wreath is built (t7).  After
day 10,700 the dynamo emerges from this state and 
returns to building two wreaths of opposite polarity which flip in
their sense an additional three times as the simulation continues.

\section{Strange States and Wreaths of a Single Polarity}
\label{sec:D5_strange_states}

\begin{figure}
  \begin{center}
  \includegraphics[width=\linewidth]{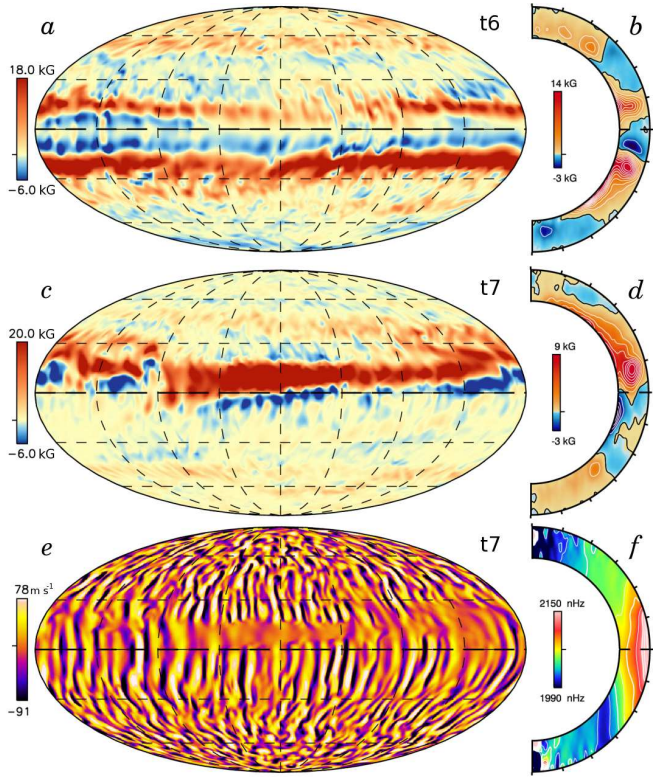}
 \end{center}
  \caption[Strange single-polarity states in case~D5]
	  {Strange single-polarity states in case~D5 as sampled at
            times~t6 and t7.  
    $(a)$~Snapshot~of~$B_\phi$ at mid-convection zone, showing two
          strong wreaths of the same polarity.   
    $(b)$~Instantaneous profile of $\langle B_\phi \rangle$ at same time.
    $(c)$~Snapshot of $B_\phi$ at mid-convection zone at a time when a
    single wreath is formed.  $(d)$~Weaker negative polarity
    structures are visible in profile of $\langle B_\phi \rangle$ at
    same instant.
    $(e)$~Accompanying snapshot of $v_r$ at mid-convection zone,
    showing flows strongly affected by magnetism.  $(f)$~The
    instantaneous differential rotation, shown here as profile of
    $\Omega(r,\theta)$, is largely unaffected by the strong wreath.
    \label{fig:case_D5_single_polarity}}  
\end{figure}

Though generally wreaths of opposite polarity are built in each
hemisphere, these oscillating dynamos occasionally wander into distinctly
different states.  This occurs for case~D5 around day 7300.
Instead of the two nearly antisymmetric wreaths of opposite polarity
above and below the equator, the dynamo enters a state where the
polarity in each hemisphere is the same, as shown in
Figures~\ref{fig:case_D5_single_polarity}$a,b$ at day 8903.  Here two
wreaths of same polarity occupy the two hemispheres and persist for
an interval of more than 500 days.  The positive-polarity $B_\phi$
reaches average amplitudes of 18~kG while the weaker negative polarity
structures have average amplitudes of only about 3~kG.  The
azimuthally-averaged profiles of $\langle B_\phi \rangle$ emphasize
that these wreaths span the convection zone and have the same
polarity everywhere.  During this interval of time, the mean poloidal
field is predominantly quadrupolar.

The dynamo can also achieve states where only a single wreath
is built in the equatorial regions, as in
Figures~\ref{fig:case_D5_single_polarity}$c,d$ at day 9590.  Here a
single strong wreath of positive polarity fills the northern
hemisphere, with $\langle B_\phi \rangle$ reaching a peak amplitude of
+18~kG.
This unique structure persists for about 800 days before
the dynamo flips polarity and builds a strong wreath of negative
polarity.  The predecessor of this new wreath can be seen in  
profiles of $\langle B_\phi \rangle$ where a much weaker
structure of negative polarity is visible in the lower convection zone.

The strong magnetic fields realized in the single wreath states react
back on the convective flows.  This is evident in the
accompanying snapshot of radial velocities at mid-convection zone
(Figure~\ref{fig:case_D5_single_polarity}$e$).  In a narrow band
spanning $0-20^\circ$ latitude and coinciding with the strong tube,
the upflows and downflows have been virtually erased.
Fluctuations in $v_\phi$ and $v_\theta$ are also very small in this
region, and the flow is dominated by the streaming flows of
differential rotation.  Within the wreath the total magnetic
energy (ME) at mid-convection zone is locally about $10~\text{to}~100$
times larger than the kinetic energy (KE), while outside the
wreath KE exceeds ME by factors of roughly $10~\text{to}~10^{4}$
at this depth. We see similar restriction of the convective
flows whenever the magnetic fields become this strong.  

The differential rotation itself
(Figure~\ref{fig:case_D5_single_polarity}$f$) is largely
unaffected by the presence of the strong magnetic wreath.  There is no
clear signature of faster flow down the middle of the wreath.
Likewise, there is little sign of the structure in profiles of the
thermodynamic variables $P, T, S,$ or $\rho$, with the mean profile
instead dominated by latitudinal variations consistent with 
thermal wind balance.

\section{Conclusions}
\label{sec:conclusions}

In this paper we have explored dynamo action in a solar-type star
rotating five times faster than our Sun currently does.  We find that
strong dynamo action can occur in the bulk of the convection
zone in this rapidly rotating sun and that the resulting magnetic
fields are organized on global scales into large wreaths.  Generally,
these wreaths have opposite polarity in each hemisphere.  They are not
isolated flux surfaces and instead are intricately linked across the
equator and to the polar regions.

This dynamo shows rich time variation, undergoing global-scale
magnetic polarity reversals roughly every 1500 days.  During a
reversal, the magnetic wreaths near the equator propagate towards the
poles and are replaced by new wreaths with opposite polarity.  
Reversals occur in both the mean toroidal field $\langle B_\phi
\rangle$ and the mean poloidal field.  The phasing between the two
global-scale fields changes with depth in the convection zone, with
the toroidal field changing first in the lower convection zone but the
poloidal field reversing first near the surface.

We have analyzed the mechanisms by which the global-scale fields are
maintained and destroyed in this dynamo.  Generally, we find that the
toroidal fields are built primarily by the $\Omega$-effect
$P_\mathrm{MS}$, where differential rotation stretches the
global-scale poloidal magnetic field into mean 
toroidal field.  This generation term generally is well correlated
with $\langle B_\phi \rangle$, but during a reversal $P_\mathrm{MS}$
appears to reverse in sign before the mean fields themselves do.
The mean poloidal field is built by the fluctuating
emf $E_\mathrm{FI}$.  This mean contribution arises
from correlations in the fluctuating velocities and magnetic fields.
Generally, as we found in \mbox{Paper I}, $E_\mathrm{FI}$ does not
appear to be well represented by a simple $\alpha$-effect where
$E_\mathrm{FI}$ is given by $\alpha \langle B_\phi \rangle$.  Rather,
$E_\mathrm{FI}$ peaks on the poleward edge of the wreaths. 

During a reversal, the wreaths from the previous cycle propagate
towards the poles.  This appears to primarily reflect a dynamo wave, with
a poleward-slip instability probably contributing to the dynamics as well.  
The spatial offset between the primary poloidal and toroidal
generation terms,  $E_\mathrm{FI}$ and $P_\mathrm{MS}$ respectively,
likely contributes to the poleward propagating dynamo wave behavior
which we see here. 

Accompanying the poleward propagating magnetic structures are changes
in the global-scale differential rotation. Bands of prograde rotating
fluid move poleward along with the magnetic 
wreaths.  These speedup bands appear to be accelerated primarily by
magnetic torques arising from the global-scale axisymmetric magnetic
fields, with an additional contribution arising from poloidal
circulations associated with the propagating magnetic wreaths.  
The angular velocity contrast $\Delta \Omega_\mathrm{lat}$
can vary by more than 45\% during a reversal and these variations may
represent changes of several percent relative to the rotating
reference frame.  As such, we may expect similar variations in
observations of surface differential rotation on magnetically active
stars during the course of a magnetic cycle.
The bands of speedup bear some resemblance to the poleward branch of
the torsional oscillations seen in observations of the Sun over the
course of a solar cycle \citep{Thompson_et_al_2003}.  In these
simulations, we do not see any evidence of the observed equatorial branch.
In the Sun, the equatorial branch may arise from enhanced cooling in
the magnetically active regions \citep[e.g.,][]{Spruit_2003, Rempel_2006, Rempel_2007},
but those effects lie beyond the spatial resolution or physical
effects included in our simulations of global-scale convection and
dynamo action.

These global-scale polarity reversals are not special to case~D5. 
Indeed, we have explored a broader class of oscillating dynamo
solutions, which will be detailed in forthcoming papers.
Some of these solutions  are realized by taking our more slowly
rotating case~D3 (\mbox{Paper I}) to higher levels of turbulence by
reducing the eddy diffusivities, while others are achieved in
simulations spinning at even higher rotation rates than case~D5.  
Large oscillations in the magnetic fields and global-scale reversals
of polarity appear to be common features in the parameter 
space we have explored, likely arising when the ohmic diffusivity on
large scales is small enough to allow the dynamo waves to run
polewards in these wreath-building dynamos.  
We find such global-scale oscillations and polarity reversals
fascinating, since these are self-consistent 3D
stellar dynamo simulations which achieve such temporally organized
behavior in the bulk of the convection zone, without appealing to a
stable tachocline of shear at the base of the convection zone as the
organizational seat of the dynamo.


\acknowledgements
The authors thank Ellen Zweibel for inspiring conversations about
dynamo theory.  This research is supported by NASA through
Heliophysics Theory Program grants NNG05G124G and NNX08AI57G, with
additional support for Brown through the NASA GSRP program by award
number NNG05GN08H and NSF Astronomy and Astrophysics postdoctoral
fellowship AST 09-02004.  CMSO is supported by NSF grant PHY 08-21899.
Miesch is supported by NASA SR\&T grant NNH09AK14I.  NCAR is
sponsored by the National Science Foundation.  Browning is supported
by the Jeffrey L.\ Bishop fellowship at CITA. Brun is partly
supported by both the Programmes Nationaux Soleil-Terre and Physique
Stellaire of CNRS/INSU (France), and by the STARS2 grant 207430 from
the European Research Council.  The simulations were carried out with
NSF PACI support of PSC, SDSC, TACC and NCSA, and by NASA HEC support
at Project Columbia.  Field line tracings shown in Figure~\ref{fig:D5
  wreaths} were produced using VAPOR \citep{Clyne_et_al_2007}.

\bibliographystyle{apj}

\end{document}